\documentclass[nofootinbib,twocolumn,superscriptaddress,a4paper]{revtex4-1}
 
\usepackage{graphicx}
\usepackage{bm,bbm}
\usepackage{xcolor}
\usepackage{tcolorbox}
\usepackage{algorithm}
\usepackage{algpseudocode}
\usepackage{amsmath}
\usepackage{amssymb}
\usepackage{textcomp}
\usepackage{booktabs} 
\usepackage[colorlinks,breaklinks]{hyperref}
\usepackage{threeparttable}
\usepackage{xr}  
\makeatletter
\newcommand\org@hypertarget{}
\let\org@hypertarget\hypertarget
\renewcommand\hypertarget[2]{%
\Hy@raisedlink{\org@hypertarget{#1}{}}#2%
  }
\makeatother
\newcommand{\ket}[1]{\left\vert#1\right\rangle}

\begin{document}


\title{Efficient integrated quantum memory for light} 
\author{Ruo-Ran Meng}
\email{These authors contributed equally to this work.}
\author{Pei-Xi Liu}
\email{These authors contributed equally to this work.}
\author{Xiao Liu}
\email{These authors contributed equally to this work.}
\author{Tian-Xiang Zhu}
\email{These authors contributed equally to this work.}
\affiliation{CAS Key Laboratory of Quantum Information, University of Science and Technology of China, Hefei 230026, China}
\affiliation{Anhui Province Key Laboratory of Quantum Network, University of Science and Technology of China, Hefei 230026, China}
\affiliation{CAS Center For Excellence in Quantum Information and Quantum Physics, University of Science and Technology of China, Hefei 230026, China}
\author{Peng-Jun Liang}
\email{These authors contributed equally to this work.}
\affiliation{CAS Key Laboratory of Quantum Information, University of Science and Technology of China, Hefei 230026, China}
\affiliation{Anhui Province Key Laboratory of Quantum Network, University of Science and Technology of China, Hefei 230026, China}
\affiliation{CAS Center For Excellence in Quantum Information and Quantum Physics, University of Science and Technology of China, Hefei 230026, China}
\affiliation{Hefei National Laboratory, University of Science and Technology of China, Hefei 230088, China}
\author{Chao Zhang}
\email{These authors contributed equally to this work.}
\affiliation{CAS Key Laboratory of Quantum Information, University of Science and Technology of China, Hefei 230026, China}
\affiliation{Anhui Province Key Laboratory of Quantum Network, University of Science and Technology of China, Hefei 230026, China}
\affiliation{CAS Center For Excellence in Quantum Information and Quantum Physics, University of Science and Technology of China, Hefei 230026, China}
\affiliation{Hefei National Laboratory, University of Science and Technology of China, Hefei 230088, China}
\author{Zhong-Yang Tang}
\author{Hong-Zhe Zhang}
\affiliation{CAS Key Laboratory of Quantum Information, University of Science and Technology of China, Hefei 230026, China}
\affiliation{Anhui Province Key Laboratory of Quantum Network, University of Science and Technology of China, Hefei 230026, China}
\affiliation{CAS Center For Excellence in Quantum Information and Quantum Physics, University of Science and Technology of China, Hefei 230026, China}
\author{Jin-Ming Cui}
\affiliation{CAS Key Laboratory of Quantum Information, University of Science and Technology of China, Hefei 230026, China}
\affiliation{Anhui Province Key Laboratory of Quantum Network, University of Science and Technology of China, Hefei 230026, China}
\affiliation{CAS Center For Excellence in Quantum Information and Quantum Physics, University of Science and Technology of China, Hefei 230026, China}
\affiliation{Hefei National Laboratory, University of Science and Technology of China, Hefei 230088, China}
\author{Ming Jin}
\email{cattom@ustc.edu.cn}
\affiliation{CAS Key Laboratory of Quantum Information, University of Science and Technology of China, Hefei 230026, China}
\affiliation{Anhui Province Key Laboratory of Quantum Network, University of Science and Technology of China, Hefei 230026, China}
\affiliation{CAS Center For Excellence in Quantum Information and Quantum Physics, University of Science and Technology of China, Hefei 230026, China}
\author{Zong-Quan Zhou}
\email{zq\_zhou@ustc.edu.cn}
\author{Chuan-Feng Li}
\email{cfli@ustc.edu.cn}
\author{Guang-Can Guo}
\affiliation{CAS Key Laboratory of Quantum Information, University of Science and Technology of China, Hefei 230026, China}
\affiliation{Anhui Province Key Laboratory of Quantum Network, University of Science and Technology of China, Hefei 230026, China}
\affiliation{CAS Center For Excellence in Quantum Information and Quantum Physics, University of Science and Technology of China, Hefei 230026, China}
\affiliation{Hefei National Laboratory, University of Science and Technology of China, Hefei 230088, China}
\date{\today}
\maketitle

\begin{figure*}[tbph]
\includegraphics [width=0.9\textwidth]{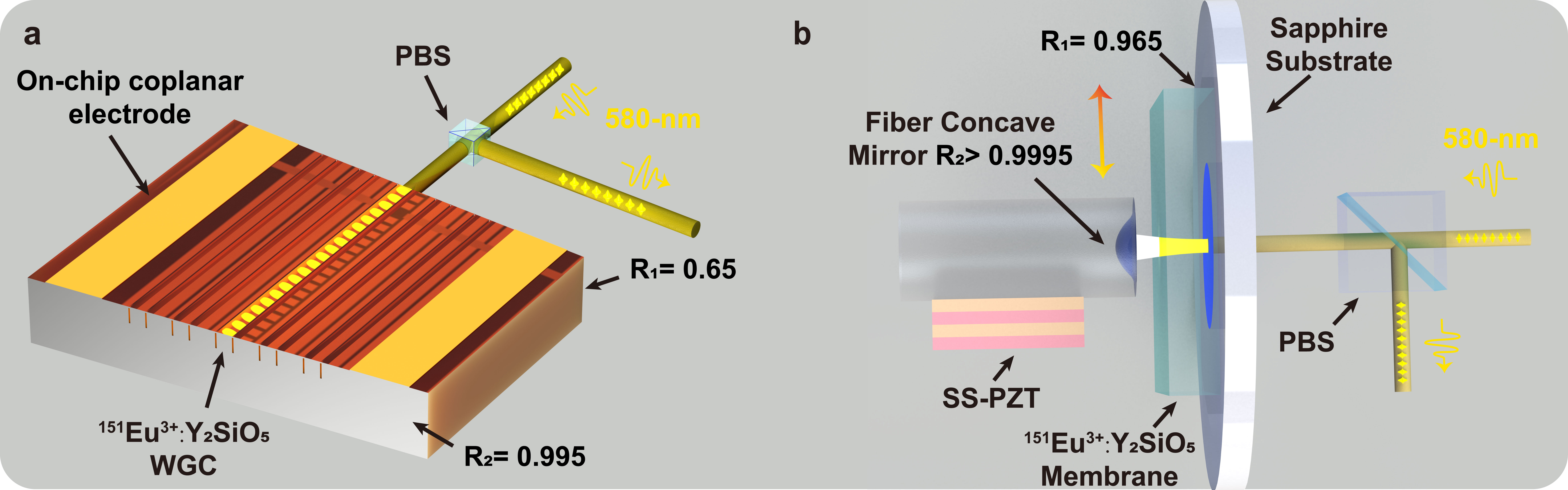}
\caption{The devices for efficient integrated quantum memories based on rare-earth-ion doped crystals. \textbf{a,} Efficient quantum memories based on the waveguide cavity (WGC) fabricated on $^{151}$Eu$^{3+}$:$\mathrm {Y_2SiO_5}$ crystals. Laser-written optical waveguides are coated with reflective layers at both ends to form optical cavities. On-chip coplanar electrodes deliver electrical pulses, enabling active control over the readout times of stored photons. Input and readout photons are separated by a polarizing beam splitter (PBS) depending on their different polarization states, with multiplexing achieved in the time domain. \textbf{b,} Efficient quantum memories based on the fiber microcavity (FBC). This cavity consists of a fiber concave mirror and a 200-\textmu m-thin membrane $^{151}$Eu$^{3+}$:$\mathrm {Y_2SiO_5}$ crystals with reflective coatings on its outer surface. The membrane is bonded to a sapphire substrate with a hole. The concave fiber mirror fixed on the stacked shearing piezoelectric transducer (SS-PZT) maintains a distance of approximately 100 \textmu m from the membrane surface. The input and output signal photons are also separated using a PBS.
}
\label{fig:FIG1}
\end{figure*}

\begin{figure*}[tbph]
\includegraphics [width=0.9\textwidth]{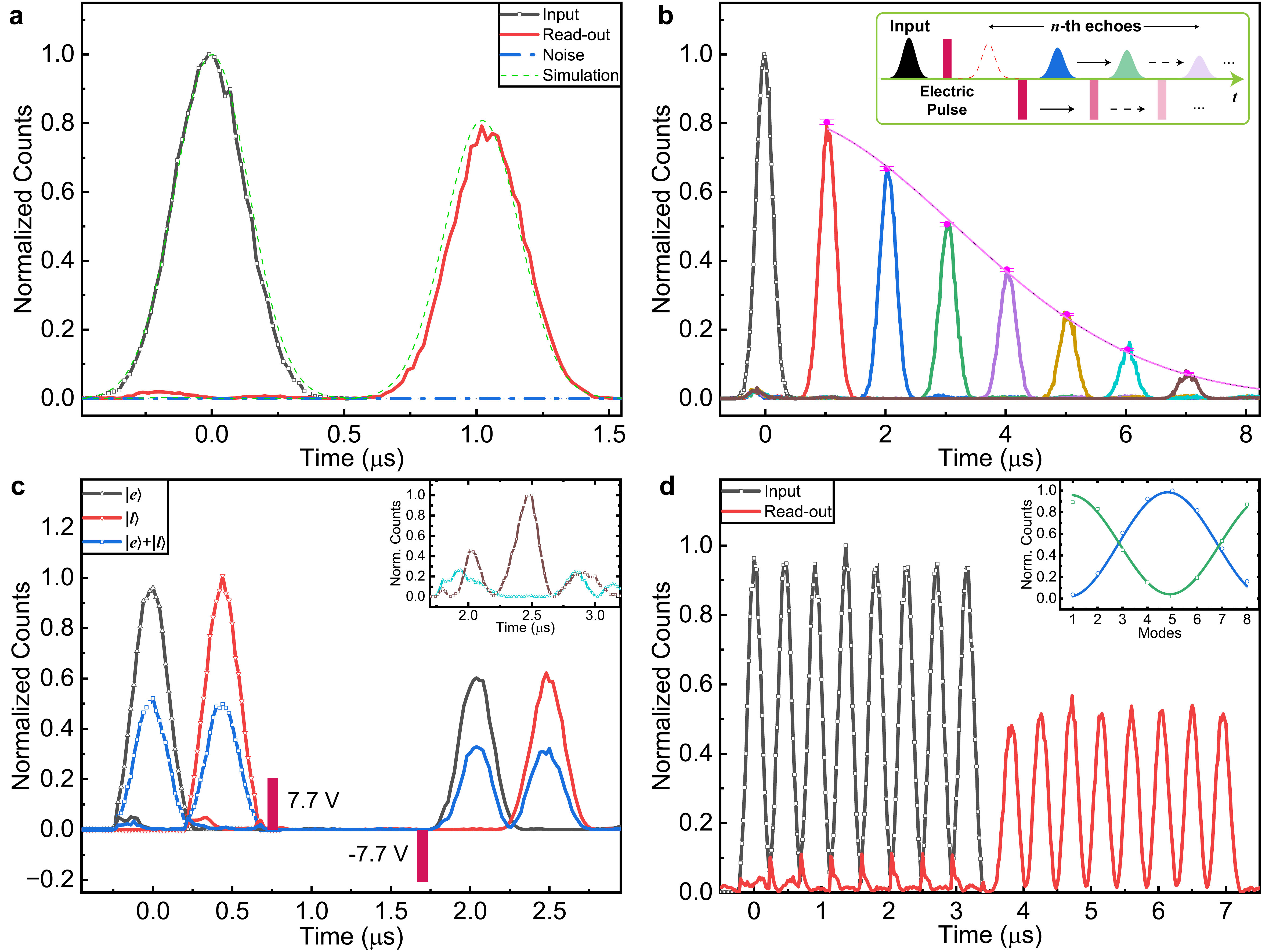}
\caption{Efficient and multiplexed single-photon-level storage in the WGC. \textbf{a,} Normalized photon counting histograms for the 1st AFC echo. The red solid line represents the 1st AFC echo, compared to the reference input signal (black solid line with hollow squares) at an average photon number per pulse $\mu=0.35$. Noise measurements with zero input are shown as a blue dot-dashed line. The storage efficiency of the 1st AFC echo is $80.3(7)\%$, while the simulation (green dashed line) predicts a slightly higher efficiency of $80.9\%$.
\textbf{b,} Photon counting histograms of higher-order AFC echoes with $\mu=0.35$. Two electric pulses actively control the readout times through Stark-shift induced interference. The fitted efficiency decay (pink solid line) yields an AFC finesse of $F_{\mathrm{AFC}}$ of 11.9(1), consistent with the expected value for the impedance-matched condition.
\textbf{c,} On-demand storage of time-bin qubits. Three input states $\ket{e}$, $\ket{l}$, and $\ket{e}+\ket{l}$ are presented in the histogram. The average photon number per qubit $\mu_q = 0.35$, with an average storage efficiency of 62.0(6)\%. The inset shows interference measurements on the superposition state $\ket{e}+\ket{l}$, with constructive (brown solid line with hollow squares) and destructive interference (cyan solid line with hollow triangles).
\textbf{d,} Storage of 8 temporal modes with an average efficiency of 52.9(2)\%. The FWHM of the temporal modes is set as $0.23$ \textmu s with each mode containing $\mu=0.17$. The inset provides the interference between the retrieved multimode echoes and reference pulses with two initial phases differing by $\pi$, yielding an average visibility of $0.94(4)$.}
\label{fig:wg}
\end{figure*}

\begin{figure*}[tbph]
\includegraphics [width=0.9\textwidth]{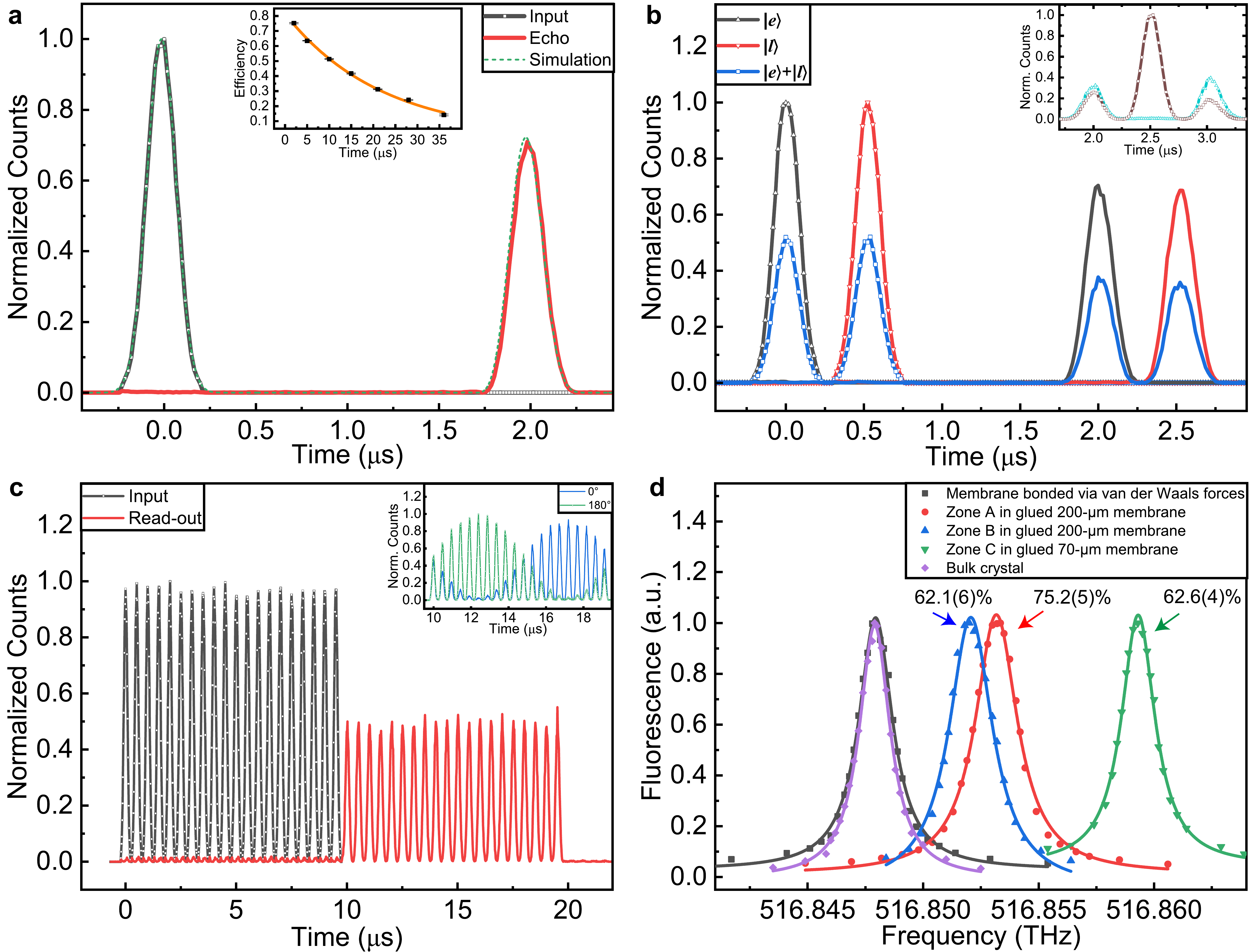}
\caption{Efficient and spectrally-tunable quantum storage using $^{151}\mathrm {{Eu}^{3+}}$:$\mathrm {Y_2SiO_5}$ membranes coupled into the FBC. \textbf{a,} Photon counting histograms of storage for weak coherent pulses with $\mu=0.35$, with an efficiency of 75.2(5)\% at 2 \textmu s, while the simulation (green dashed line) predicts a slightly higher efficiency of $78.4\%$. The inset shows efficiency decay fitted with an exponential decay function $\mathrm{exp}(-4 t/T_{\mathrm{AFC}}^{\mathrm{eff}})$, yielding an effective AFC lifetime $T_{\mathrm{AFC}}^{\mathrm{eff}}$ = 87.6(2.4) \textmu s.   
\textbf{b,} Storage of time-bin qubits with $\mu_q=0.7$. The inset shows the interference measurements for the superposition state $\ket{e}+\ket{l}$.
\textbf{c,} Quantum storage of 20 temporal modes with average efficiency of 51.3(2)\% at $\mu=0.7$. The inset shows the interference between retrieved multimode echoes and reference pulses, yielding an average visibility of 0.94(1).
\textbf{d,} Photoluminescence excitation spectra of $^{151}\mathrm {{Eu}^{3+}}$:$\mathrm {Y_2SiO_5}$ membranes under variable strain. Lorentz fits reveal fluorescence centers at 516.8479 THz (black), 516.8479 THz (purple), 516.8521 THz (blue), 516.8532 THz (red), and 516.8593 THz (green). Quantum storage efficiencies achieved at the latter three frequencies are 62.1(6)\%, 75.2(5)\%, and 61.6(4)\%, respectively.}
\label{fig:storage}
\end{figure*}

\begin{figure*}[tbph]
\includegraphics [width=1\textwidth]{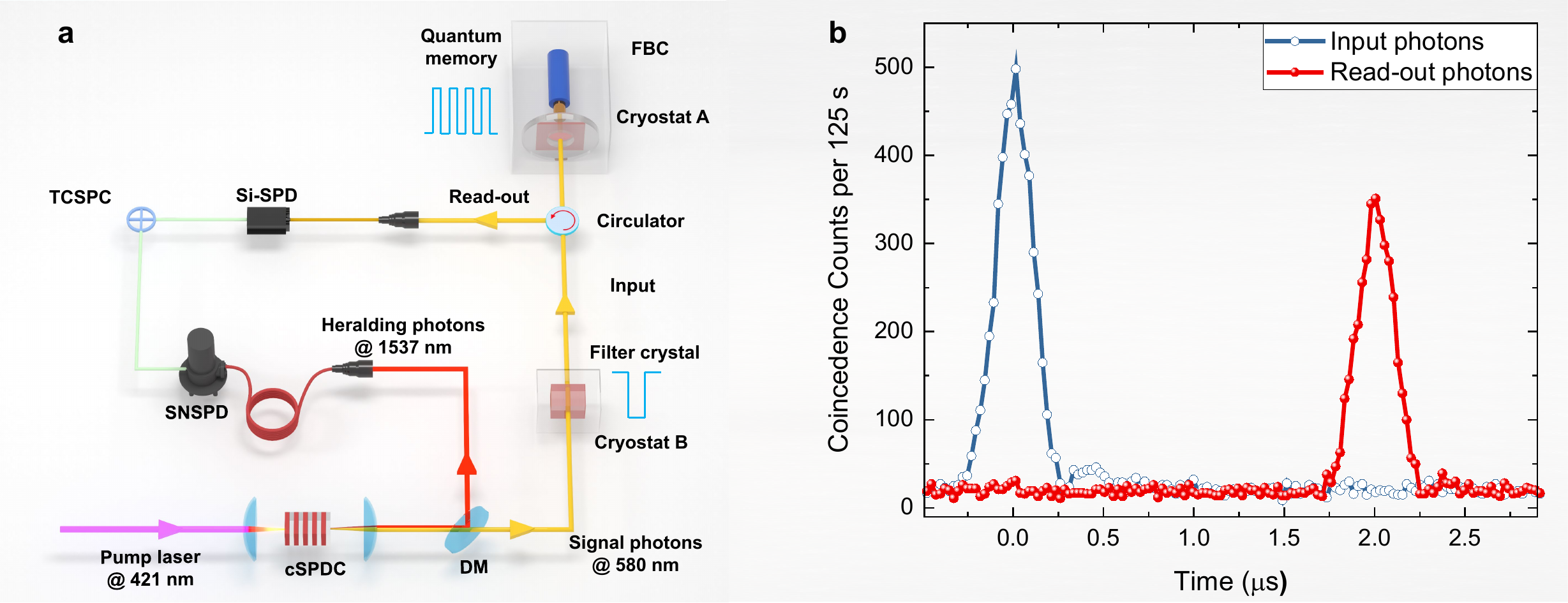}
\caption{Efficient storage of telecom-heralded single photons in the FBC. \textbf{a,} A 421-nm laser pumps a nonlinear crystal, generating photon pairs at 1537 nm and 580 nm via cavity-enhanced spontaneous parametric down-conversion (cSPDC). Photon pairs are separated by a dichroic mirror (DM). The 1537-nm telecom-band photons are detected by a superconducting nanowire single-photon detector (SNSPD), providing heralding signals. The 580-nm signal photons are spectrally filtered using an optically-pumped filter crystal, resulting in a bandwidth of 4 MHz. After storage in the fiber microcavity (FBC) quantum memory, the readout photons are extracted by an optical circulator and detected by a silicon-based single-photon detector (Si-SPD). Coincidence measurements of the photon pairs are performed using a time-correlated single-photon counting (TCSPC) module. \textbf{b,} Photon counting histograms for storage of heralded single photons. The measured storage efficiency is 69.8(1.6)\% at 2-\textmu s with a cross-correlation of 16.4(2) within a 123-ns window, confirming non-classical correlations between retrieved photons and telecom heralding photons.}
\label{fig:singlephoton}
\end{figure*}

\textbf{
Scalable implementation of quantum networks and photonic processors demands integrated photonic memories with high efficiency, yet current integrated systems have been limited to storage efficiencies below 27.8\%. Here, we demonstrate highly efficient integrated quantum memories based on rare-earth-ion-doped crystals coupled with impedance-matched microcavities, realized in two novel architectures: 200-\textmu m-thin membranes of $\mathrm {{Eu}^{3+}}$:$\mathrm {Y_2SiO_5}$ integrated with fiber-based microcavities, and waveguide-based cavities fabricated using femtosecond lasers. Our approach achieves reliable integrated quantum storage with record efficiencies of 80.3(7)\% for weak coherent pulses and 69.8(1.6)\% for telecom-heralded single photons, alongside the storage of 20 temporal modes with an average efficiency of 51.3(2)\%. Moreover, the thin-membrane $\mathrm{{Eu}^{3+}}$:$\mathrm{Y_2SiO_5}$ architecture enables spectrally tunable efficient quantum storage via variable strain, providing a flexible interface for quantum networks. By combining high efficiency, large multimode capacity, and tunability, our devices establish a versatile hardware foundation for scalable quantum repeaters and chip-scale photonic processors. 
}


Optical quantum memory, capable of storing and retrieving quantum states of light, is a foundational technology for long-distance quantum networks \cite{RevModPhys.83.33,Ritter2012An,Liu2021} and for photonic information processing \cite{RevModPhys.79.135,PhysRevLett.110.133601,liu2024nonlocal}. Storage efficiency is a key figure of merit \cite{RevModPhys.83.33,Zhou2023Photonic} because higher efficiency can boost the rate of entanglement distribution in quantum repeaters \cite{RevModPhys.83.33} and improves the success probabilities of multiphoton generation and quantum gate operations in optical quantum computing \cite{RevModPhys.79.135,liu2024nonlocal,PhysRevLett.110.133601}. A pivotal efficiency threshold is 50\%, which is the necessary condition to operate in the no-cloning regime without post-selection \cite{PhysRevA.64.010301} and for implementing error correction protocols in one-way quantum computation \cite{PhysRevLett.97.120501}.

Quantum storage of light leverages collective enhancement in atomic ensembles, whether gaseous or solid-state \cite{Julsgaard2004,Wang2019,Hosseini2011,deRiedmatten2008,Hedges2010,England2013From}, or strong coupling of single atoms with optical cavities \cite{specht2011single,Korber2018Decoherence}. To date, the highest reported quantum storage efficiency has reached 90.6\% in cold atomic gases using the electromagnetically induced transparency scheme \cite{Vernaz-Gris2018,Wang2019}, 82\% in warm atomic vapors via Raman memory \cite{Guo2019}, and 69\% in rare-earth ions (REI)-doped solids employing the gradient echo memory scheme \cite{Hedges2010}, all operating on micro-second timescales. However, these approaches require physically large medium (10$^{-1}$$\sim$10$^{4}$ mm$^3$) and are limited to single temporal modes because the underlying memory protocols suffer from inherently low multimode capacity \cite{PhysRevLett.101.260502}. Integrated quantum memories for light have been demonstrated in REI-doped solids \cite{Saglamyurek2011,Zhong2017Nanophotonic,Liu2020On-demand,Zhou2023Photonic,PhysRevLett.113.053603,Seri2019Quantum}, successfully reducing the medium size by several orders of magnitude (Fig.~\ref{fig:comparison}), but the highest storage efficiency achieved so far remains limited to 27.8\% \cite{Liu2020On-demand}, presenting a significant hurdle for their practical deployment in quantum networks.

In this work, we introduce highly efficient integrated quantum memories based on $^{151}\mathrm {{Eu}^{3+}}$:$\mathrm {Y_2SiO_5}$ crystals coupled with impedance-matched optical cavities. Two novel microcavity configurations are utilized for the quantum storage of light: the waveguide cavity (WGC) and the fiber-based microcavity (FBC), as shown in Fig. \ref{fig:FIG1}. The WGC is constructed by applying reflective coatings to the end surfaces of an on-chip laser-written optical waveguide, incorporating coplanar electrodes for on-demand retrieval via electric pulse control \cite{Liu2020On-demand}. This configuration achieves a storage efficiency of 80.3(7)\% for single-photon-level weak coherent inputs, the highest efficiency reported for photonic storage in solid-state systems. The FBC employed hundred-micron $^{151}\mathrm {{Eu}^{3+}}$:$\mathrm {Y_2SiO_5}$ membranes integrated into a fiber-based microcavity, achieving a storage efficiency of 69.8(1.6)\% for telecom-heralded single photons. It successfully stores 20 temporal modes with an average efficiency exceeding 50\%. Additionally, by applying strain to the membranes, the optical absorption center is tuned over a 10 GHz range, enabling efficient quantum storage with adjustable center frequencies. While microcavities have previously been exploited to address individual rare-earth ions in the Purcell regime \cite{PhysRevX.10.041025,Casabone2021}, our FBC operates in a distinct regime that harnesses the collective coupling of a rare-earth-ion ensemble with light.

\section{Results}
\subsection{Theoretical description}
Efficient storage of light intuitively demands a medium with strong optical absorption. However, the coherent intra-$4f$ optical transitions of REI are generally weak due to the parity-forbidden nature of the electric dipole transition. This challenge is particularly prominent for the $^{7}F_{0}$ $\rightarrow$ $^{5}D_{0}$ optical transition of the $\mathrm {{Eu}^{3+}}$:$\mathrm {Y_2SiO_5}$ crystal, a promising candidate for long-lived quantum memory that has enabled coherent light storage for 1 hour \cite{Zhong2015,Ma2021}. 
Preliminary approaches have been explored to achieve efficient integrated quantum memories, such as increasing the interaction length with long bent or spiral on-chip waveguides, or integrating quantum memories with photon sources in a single cavity \cite{Lau2025Efficient}; however, these approaches have yet to reach high efficiency and remain technologically demanding. A broadly applicable solution to address the weak absorption issue is to introduce an impedance-matched optical cavity, which enables near-unity storage efficiency even with optically thin medium \cite{FLEISCHHAUER2000395,PhysRevA.82.022311,Afzelius2010Impedance-matched}. 
The impedance-matched condition is given by $R_{1}=e^{-2\widetilde{d}}$ and $R_{2}=1$, where $R_{1}$ and $R_{2}$ are the reflectivities of the coupling and second mirrors, respectively, and $\widetilde{d}$ is the effective absorption depth averaged over the input bandwidth. Under this condition, assuming $\epsilon  \ll \widetilde{d}  \ll 1$, the storage efficiency of a cavity-enhanced quantum memory simplifies to \cite{PhysRevA.82.022311,Afzelius2010Impedance-matched}: 
\begin{eqnarray}
{\eta = \frac{\eta_{\mathrm{M}} \cdot \eta_{\mathrm{deph}}}{(1+\frac{\epsilon}{4\widetilde{d}})^4}},
\label{efficiency}
\end{eqnarray}
where $\epsilon$ is the optical depth of round-trip intra-cavity loss, $\eta_{\mathrm{M}}$ is the mode-matching efficiency between the input light and the cavity mode, and $\eta_{\mathrm{deph}}$ accounts for the remaining atomic dephasing. Achieving efficiencies close to 100\% requires $\eta_{\mathrm{M}}, \eta_{\mathrm{deph}}\approx 1$ and $\epsilon$/${4\widetilde{d}} \ll 1$.

This impedance-matched cavity approach has been
implemented with bulk REI-doped crystals using the atomic frequency comb (AFC) protocol \cite{PhysRevLett.110.133604,Jobez_2014,Duranti2024Efficient}, whose absorption profile consists of a series of periodic peaks. To date, however, the efficiencies achieved remain below that of single-pass long-crystal configurations \cite{Hedges2010}. Here we present two compact cavity designs that simultaneously deliver high $\eta_{\mathrm{M}}$ and low $\epsilon$, enabling efficient integrated quantum storage.
For cavity-enhanced AFC memories, achieving $\eta_{\mathrm{deph}} \approx 1$ requires a high finesse of the AFC structure($F_{\mathrm{AFC}}$), which places stringent demands on the peak absorption depth $d$, as $F_{\mathrm{AFC}} \approx d$/$\widetilde{d}$ \cite{Jobez_2014}. However, creating an AFC with strong absorption often induces a slow-light effect \cite{PhysRevLett.111.183602,Duranti2024Efficient}, which limits the achievable bandwidth and hinders temporal multimode storage. To solve this dilemma, we employ an absorption-enhanced spectral hole-burning technique to prepare AFC with enhanced absorption embedded within a transparency pit, so that the fast-light dispersion caused by the enhanced AFC could counteract the slow-light dispersion caused by the spectral pit (See Supplementary Information for details). This approach enables the preparation of high-finesse AFC while suppressing slow-light effects, thereby preserving a relatively large system bandwidth and facilitating efficient storage of multiple temporal modes.

\begin{table*}

    \caption{\label{TABLE 2} \textbf{The device parameters for FBC and WGC} 
     }
    \begin{ruledtabular}
    \begin{threeparttable}
    \begin{tabular}{ccccccccc}
        System &
        $\eta_{\mathrm{M}}$ & 
        $F_{\mathrm{AFC}}$ &
        $\eta_{\mathrm{deph}}$&
        $\epsilon$/${4\widetilde{d}}$ & 
        $(1+\epsilon$/$4\widetilde{d})^{-4}$ & 
        $\eta_{\mathrm{Theo}}$ & 
        $\eta_{\mathrm{Exp}}$ 
        \\ 
        \hline
        FBC &
        $95\%$ &
        $8$ &
        $89.6\%$ & 
        $1/40$ &
        $90.6\%$ &
        $77.1\%$ &
        $75.2(5)\%$\\

        WGC &
        $98.5\%$ &
        $12$ &
        $95.2\%$ & 
        $1/28$ &
        $86.9\%$ &
        $81.5\%$ &
        $80.3(7)\%$\\

    \end{tabular}
     \begin{tablenotes}
        \footnotesize
        \item 
        $\eta_{\mathrm{Theo}}$: Theoretical storage efficiency derived from Eq. \ref{efficiency}.
        $\eta_{\mathrm{Exp}}$: The experimentally measured storage efficiency.
        
      \end{tablenotes}
  \end{threeparttable}
   
    \end{ruledtabular}

\end{table*}

\subsection{Memory Performance}
Gaussian-shaped weak coherent pulses with a full width at half maximum (FWHM) of 0.45 \textmu s and 0.19 \textmu s are used for the WGC and FBC, respectively, with average photon numbers per pulse ($\mu$) of 0.35 and 0.70. As shown in Fig.~\ref{fig:wg}a and~\ref{fig:storage}a, single pulses are stored and retrieved for 1 \textmu s for the WGC and 2 \textmu s for the FBC, achieving efficiencies of $80.3(7)\%$ and $75.2(5)\%$, respectively. Although FBC exhibits a lower $\epsilon$/${4\widetilde{d}}$ than WGC, the WGC achieves a higher storage efficiency due to its higher $\eta_{\mathrm{M}}$ and $\eta_{\mathrm{deph}}$, as summarized in Table \ref{TABLE 2}. The quantum storage efficiency of WGC surpasses both the highest quantum storage efficiency (69\%) \cite{Hedges2010} and the highest light storage efficiency (76\%) \cite{PhysRevLett.116.073602} achieved in any solid-state medium. The theoretical efficiency of the WGC is estimated to be 81.5\% according to Eq. \ref{efficiency} and the simulated efficiency based on measured cavity parameters is 80.9\% (Fig.~\ref{fig:wg}a). Both numbers align well with our experimental results. Details about the simulation of the storage process are provided in the Supplementary Information. Reducing $\epsilon$ from 0.029 to 0.01 in the WGC could enable storage efficiency of above 90\%.

We demonstrate on-demand retrieval of input photons in the WGC using Stark-shift-induced interference for active control of readout times \cite{Horvath2021Noise-free,Liu2020On-demand}. After the absorption of input photons, two electric pulses (FWHM of 48 ns, voltage of $\pm$7.7 V) are applied to read out the photons in an arbitrary order of AFC echoes. Fig.~\ref{fig:wg}b shows the 1st to 7th AFC echoes with $\mu=0.35$, where efficiency decay follows a Gaussian profile, yielding a fitted $F_{\mathrm{AFC}}$ of 11.9(1), consistent with the expected value for the impedance-matched condition.

To benchmark the quantum storage fidelity, time-bin qubits are encoded onto input photons. Photon-counting histogram for qubit states $\ket{e}$, $\ket{l}$ and $\ket{e}$+$\ket{l}$ are shown in Fig.~\ref{fig:wg}c and~\ref{fig:storage}b, where $\ket{e}$ and $\ket{l}$ represent the early and the late qubit states, respectively. The fidelities of the superposition states ( $\ket{e}+\ket{l}$ and $\ket{e}+\mathrm{i}\ket{l}$) are analyzed using an independent AFC-based analyzer for the FBC \cite{Duranti2024Efficient} and a double-AFC structure for the WGC \cite{Liu2020On-demand} (see Supplementary Information for details). The measured fidelities, summarized in Supplementary Table 3 in the Supplementary Information, reveal a minimum total fidelity of 99.0(1)\%, which indicates a strong violation of the fidelity bound that can be achieved with the classical measure-and-prepare strategy \cite{Liu2020On-demand,specht2011single}, confirming that the memories are unambiguously operating in the quantum regime. We note that the large bandwidth achieved in the FBC ensures that the storage efficiency for time-bin qubits (74.6(6)\%) closely matches that for single pulses (75.2(5)\%). This establishes the first solid-state quantum memory for photonic qubits with an efficiency conclusively surpassing the 50\% threshold, a critical cornerstone for practical applications in quantum networks and photonic processors.

In the FBC configuration, we extend the storage time to 10 \textmu s and demonstrate the storage of 20 temporal modes with an average storage efficiency of $51.3(2)\%$ (Fig.~\ref{fig:storage}c). For the WGC, 8 temporal modes are stored with an average storage efficiency of $52.9(2)\%$ over a storage time of 3.8 \textmu s (Fig.~\ref{fig:wg}d). The reduced temporal multimode capacity of the WGC is attributed to a narrower storage bandwidth and a shorter two-level AFC lifetime. The former is constrained by the cavity linewidth due to a stronger slow-light effect, while the latter is limited by a shorter optical coherence lifetime (measured with two-pulse photon echoes), most likely induced by stronger vibrations in the WGC cryostat.

To confirm coherence preservation in the retrieved multimode echoes, we measure the interference fringes between retrieved echoes and reference pulses (see Supplementary Information for details). Given the multiple temporal modes, we observe a complete sine oscillation fringe by linearly varying the phase of all input modes while maintaining the reference pulse phase constant (inset in Fig.~\ref{fig:wg}d and~\ref{fig:storage}c). The measured interference visibilities of 0.94(4) for the WGC and 0.94(1) for the FBC confirm reliable quantum storage of multiple temporal modes.

In the FBC configuration, the membrane flat mirror is bonded to a sapphire substrate with a central hole using epoxy resin adhesive. This bonding process introduces variable strain to the membrane as the adhesive solidifies and cools, shifting the optical absorption center frequency of $^{151}$Eu$^{3+}$ ions via the piezospectroscopic effect \cite{Galland2020Mechanical}.
As shown in Fig.~\ref{fig:storage}d, we measure the photoluminescence spectra of membranes. For the aforementioned 200-\textmu m membrane with a storage efficiency of 75.2(5)\%, the absorption center frequency is observed at 516.8532 THz. At a different location on the same membrane, the absorption center frequency shifts to 516.8521 THz, where a quantum storage efficiency of 62.1(6)\% is measured for weak coherent pulses. Furthermore, a thinner membrane with a thickness of 70 \textmu m shows an absorption center frequency of 516.8593 THz, with a measured storage efficiency of 61.6(4)\%. These variations in efficiency are attributed to increased intra-cavity loss, likely caused by dust on the fiber concave mirror during membrane switching. Notably, membranes attached to fused quartz substrates via van der Waals forces exhibit an absorption center frequency of 516.8479 THz, nearly identical to that of bulk crystals without additional strain. An optical frequency shift of 10 GHz implies a pressure of $\sim$47 MPa in $\mathrm{Eu^{3+}}$:$\mathrm{Y_2SiO_5}$ crystals \cite{Thorpe2011}. This significant pressure likely originates from shear strain generated in the thin membranes. While strain in bulk crystals typically induces both frequency shifts and additional broadening \cite{Galland2020Mechanical}, strain in hundred-micron membranes can cause substantial optical frequency shifts ($\sim$10 GHz) with negligible broadening, thus preserving the original absorption depth for efficient quantum storage. This property enables the realization of spectrally tunable and efficient quantum memories, which could facilitate wideband spectral multiplexing within a single device through controlled strain, and flexible interfacing with quantum light sources based on free atoms, where achieving large spectral tuning on the atomic side poses significant challenges.

\subsection{Storage of single photons}
All previous demonstrations of efficient solid-state quantum memories \cite{Hedges2010,Duranti2024Efficient} are implemented with attenuated laser pulses rather than true quantum light, despite the latter being essential for practical quantum network applications such as quantum repeaters \cite{Liu2021,RevModPhys.83.33,Lago-Rivera2021}, photonic quantum computing \cite{liu2024nonlocal,RevModPhys.79.135,PhysRevLett.110.133601}. To date, the maximum storage efficiency for single photons in solids has been limited to 18\% \cite{Liu2021,businger2022non,liu2024nonlocal}, far below the requirements for practical applications.
Here, we demonstrate the efficient storage of single photons with the heralding photon at the telecom C-band \cite{zhu2025metropolitan} using the setup illustrated in  Fig.~\ref{fig:singlephoton}a. Heralded single-photon generation is achieved via cavity-enhanced nondegenerate spontaneous parametric down-conversion in a periodically poled potassium titanyl phosphate (PPKTP) crystal pumped by a 421-nm laser. The PPKTP crystal is housed in a symmetric confocal cavity supporting multi-wavelength resonance, producing photon pairs at 580 nm and 1537 nm. After spectral filtering with interference filters, cascaded etalons, and a filter crystal based on an optically-pumped 1\%-doped $\mathrm {{Eu}^{3+}}$:$\mathrm {Y_2SiO_5}$ crystal, the final bandwidth of the 580-nm photons is limited to 4 MHz. These telecom-heralded 580-nm single photons are stored in the FBC-based quantum memory. We measure a storage efficiency of $(69.8\pm1.6)$\% at 2 \textmu s (see Fig.~\ref{fig:singlephoton}b), setting a new record for the storage efficiency of true single photons in solids, which also significantly outperforms the corresponding efficiency ($\sim$33\%) of a fiber delay line at the same wavelength. The slight efficiency drop relative to weak-coherent inputs arises from the spectrally filtered single-photon wave-packet departing from an ideal Gaussian profile. The second-order cross-correlations between 1537-nm photons and readout 580-nm photons are measured to be 16.4(2) within a 123-ns window, far exceeding the classical threshold of 2, confirming the non-classical nature of the retrieved signal \cite{businger2022non}. With a single-mode duration of 0.5 \textmu s for the heralded single photons, the FBC successfully stored four temporal modes of heralded single photons. Given the C-band heralding photons, this system provides a direct interface for long-distance quantum networks \cite{Liu2021,RevModPhys.83.33,zhu2025metropolitan}.

\begin{figure*}[tbph]
\includegraphics [width=0.8\textwidth]{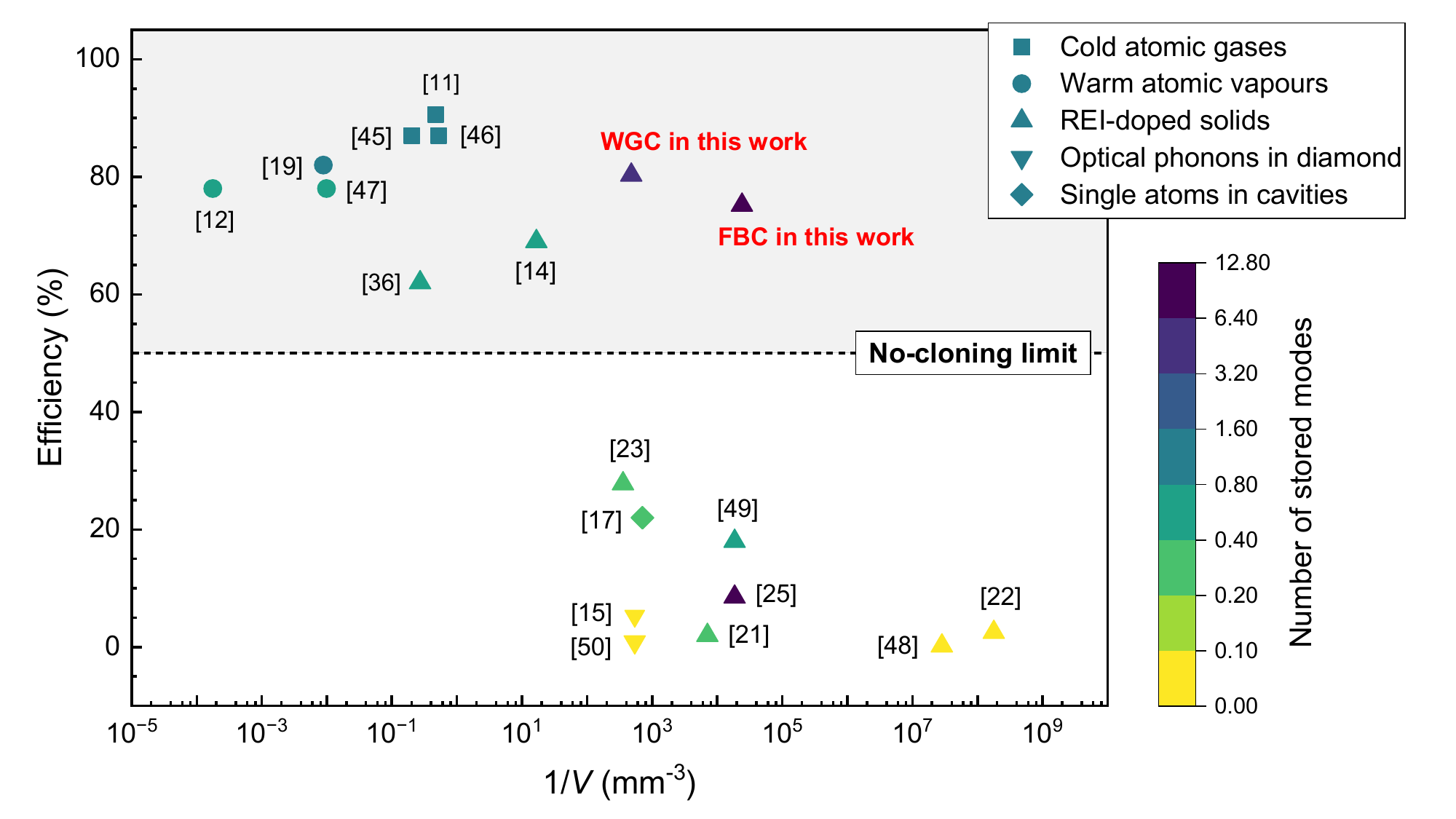}
\caption{Performance overview of quantum memories for light. The horizontal axis represents the inverse of effective volume which corresponds to the number of potential memory units per volume of mm$^3$. The surveyed physical systems include cold atomic gases \cite{Cao:20,Wang2019,Cho:16}, warm atomic vapors \cite{Guo2019,Ma2022,Hosseini2011}, rare-earth ion (REI)-doped solids \cite{Hedges2010,Duranti2024Efficient,Zhong2017Nanophotonic,Craiciu2019Nanophotonic,Rakonjac2022Storage,Liu2020On-demand,Saglamyurek2011,Seri2019Quantum}, optical phonons in diamond \cite{England2013From,England2015Storage}, and single atoms in cavities \cite{Korber2018Decoherence}, with a focus on those that have demonstrated efficient or small-volume quantum storage. For each experimental work plotted, the reported efficiency corresponds to the maximum measured storage efficiency. The effective volume $V$ of a quantum memory is determined by multiplying the transverse mode area of the optical field with the effective device length (the medium length or the cavity length). The effective number of stored modes is calculated as the product of the number of independent modes and the corresponding storage efficiency, with the color depth of each data point representing this value.}
\label{fig:comparison}
\end{figure*}

\section{Discussion}

In conclusion, we have demonstrated integrated quantum memories for light based on both the WGC and the FBC. We achieve multiplexed quantum storage with an efficiency up to 80.3(7)\% for weak coherent pulses and 69.8(1.6)\% for telecom-heralded single photons. Fig. \ref{fig:comparison} presents an overview of efficient and compact quantum memories implemented across various physical systems. This work represents the first demonstration of multimode quantum storage with an efficiency surpassing the no-cloning limit of 50\%, while achieving a device volume as small as $4\times10^{-5}$ mm$^3$, a reduction of more than three orders of magnitude compared to other efficient quantum memories. 

The FBC and the WGC offer complementary strengths in applications. The FBC architecture supports spectrally tunable and efficient quantum storage through variable strain applied to thin membranes, facilitating wideband spectral multiplexing and flexible interfacing with quantum light sources. This tunable quantum memory complements single-atom-based tunable quantum light sources \cite{Trotta2016,Huang_2023}, forming an efficient interface that could dramatically boost the data rate of multiplexed quantum repeaters \cite{Liu2021,zhu2025metropolitan}. Meanwhile, WGCs offer the potential for high-density integration via 3D FLM, enabling spatial multiplexing.

Extending the FBC to long-lived spin-wave storage is straightforward: the planar mirror coating can be engineered for high reflectance at normal incidence but low reflectance at large angles, allowing control pulses to be injected non-collinearly with respect to the signal \cite{Ma2021Elimination,Ortu2022}. For the WGC, spin-wave operation can be realized by dual-resonance of two orthogonal polarizations and subsequent polarization-based filtering of the control pulse \cite{doi:10.1126/sciadv.adu5264}. The WGC facilitates easy operation at the ``magic” magnetic field, which demands precise tuning of the entire memory device \cite{Zhong2015,Ma2021}. Moreover, fiber pig-tailing and packaging are essential to enhance the practicality of integrated quantum memories. We have previously demonstrated waveguide-based quantum memories with fiber pig-tailing \cite{PhysRevLett.129.210501}. Extending this capability to WGCs and FBCs introduces additional challenges, with efficient mode matching as a key issue that can potentially be addressed using thermally expanded core fibers or graded-index lenses.

Our results overcome the efficiency barriers of integrated quantum memories and highlight their unique advantages in high-density multiplexing and flexible spectral tunability, clearing the way for their widespread application in large-scale quantum networks and photonic processors.

\section{Methods}
\subsection{WGC configuration}
In the WGC configuration (Fig.~\ref{fig:FIG1}a), the substrate is a $^{151}\mathrm {{Eu}^{3+}}$:$\mathrm {Y_2SiO_5}$ crystal (doping level: 700 ppm) with dimensions $5.0 \times 4.0 \times 4.9$ mm$^3$ along the crystal's $D_{1} \times D_{2} \times b$ axes. The $D_{1} \times D_{2}$ facets are polished to form a 12-arcsecond wedge and coated with reflective films, providing reflectivities of $R_1$ = 65\% and $R_2$ = 99.5\% at 580 nm, respectively. On-chip electrodes fabricated along the $b$ axis, with 380 \textmu m spacing, enable on-demand retrieval via electric pulse control \cite{Liu2020On-demand}. Five Type-IV optical waveguides are fabricated between each pair of electrodes using femtosecond laser micromachining (FLM) technology. The wedge ensures the cavity resonance of at least one waveguide within the absorption profile of $^{151}\mathrm{Eu}^{3+}$. Coating directly on bulk crystals to form a cavity \cite{PhysRevLett.110.133604} significantly reduces intra-cavity losses compared to large open cavities \cite{Duranti2024Efficient}, but introduces significant spatial mode mismatching. By incorporating optical waveguides, we successfully mitigated this mismatch in the WGC, achieving a mode-matching efficiency of $\eta_{\mathrm{M}}=98.5\%$, a significant improvement over the 84\% mode-matching efficiency reported in Ref.~\cite{PhysRevLett.110.133604}. After systematic optimization of the fabrication process, the optical waveguide finally achieves ultralow propagation loss of 0.012 (0.1 dB/cm, see Supplementary Information), with an estimated $\epsilon$/${4\widetilde{d}}=1/28$.

\subsection{FBC configuration}
In the FBC configuration (Fig.~\ref{fig:FIG1}b), the $^{151}\mathrm {{Eu}^{3+}}$:$\mathrm {Y_2SiO_5}$ membrane, cut from the same crystal boule, has a size of $5 \times 4 \times 0.2$ mm$^3$ along the $D_{1} \times D_{2} \times b$ axes. The concave fiber mirror is fabricated using CO$_2$ laser machining \cite{Hunger_2010}. To minimize intra-cavity losses, the outer surface of the membrane is directly coated with reflective films, serving as the flat cavity mirror. Reflectivities for the stored light at 580 nm are $R_1$=96.5\% (membrane mirror) and $R_2$=99.9\% (fiber mirror), with an estimated $\epsilon$/${4\widetilde{d}}$ of 1/40.
Input light is coupled into the cavity via the membrane flat mirror with $\eta_{\mathrm{M}}=95(1)\%$. To mitigate cryostat vibration effects on cavity length, passive isolation using springs and active locking with shear piezoelectric actuators are employed. The locked cavity has a length stability below 50 pm, resulting in a storage efficiency loss of less than 0.2\% (see Supplementary Information for details).

\subsection{AFC configuration}
Following the absorption-enhanced spectral hole-burning process, the absorption depth $d$ of Eu$^{3+}$ ions in the target band is increased to approximately 2.6 for the WGC and 0.14 for the FBC. In the AFC preparation process, we prepare AFC with a total bandwidth of 6 MHz, with comb periodicity of 1 MHz and 500 kHz for the WGC and FBC, respectively. To optimize the impedance-matched condition, the finesse of the AFC is adjusted, yielding estimated values of 12 for the WGC and 8 for the FBC. Additional details of the spectral preparation process are provided in the Supplementary Information.


\bigskip
\textbf{Acknowledgments}
This work is supported by the Innovation Program for Quantum Science and Technology (No. 2021ZD0301200), the National Natural Science Foundation of China (Nos. 12222411, 11821404, 12474367, and 12204459), and this work is partially carried out at the USTC Center for Micro and Nanoscale Research and Fabrication. Z.-Q. Z. acknowledges the support from the Youth Innovation Promotion Association CAS.\\

\textbf{Author contributions}
Z.-Q.Z. designed the experiments and supervised all aspects of this work. M.J., R.-R.M., and X.L. performed the FBC experiments with technical supports from Z.-Y.T. and J.-M.C. M.J. locked the cavity for FBC. R.-R.M. realized quantum storage with FBC. X.L. fabricated the device for FBC. P.-X.L. and T.-X.Z. conducted the WGC experiments with assistance from H.-Z.Z. P.-X.L. performed the simulation. T.-X.Z. designed the pumping strategy. P.-J.L. grew the crystal. C.Z. constructed the quantum light source. R.-R.M., P.-X.L., X.L., T.-X.Z., M.J. and Z.-Q.Z. wrote the manuscript, with inputs from others. Z.-Q.Z. and C.-F.L. supervised the project. All authors contributed to discussing the experiments and results.\\

\textbf{Competing interests}
The authors declare no competing interests.\\

\textbf{Data availability}
The data presented in the figures within this paper and other findings of this study are available from the corresponding authors upon request.\\

\textbf{Code availability}
The custom codes used to produce the results presented in this paper are available from the corresponding authors upon request.

\clearpage

\newpage

\onecolumngrid

\renewcommand{\figurename}{Fig.}
\renewcommand{\tablename}{Table.}

\setcounter{table}{0}
\renewcommand{\thetable}{S\Roman{table}}
\setcounter{figure}{0}
\renewcommand{\thefigure}{S\arabic{figure}}
\setcounter{equation}{0}
\renewcommand{\theequation}{S\arabic{equation}}

\begin{center}{Supplementary Information for \\
"Efficient integrated quantum memory for light"}
\end{center}

\subsection*{1. Growth of $\mathrm {{Eu}^{3+}}$:$\mathrm {Y_2SiO_5}$ crystals}

The $\mathrm {{Eu}^{3+}}$:$\mathrm {Y_2SiO_5}$ crystals are grown using the Czochralski method in an argon atmosphere, with a pulling rate of 1 mm/h and a rotation rate of 25 rpm along the b-axis. High-purity SiO$_2$ (99.9999\%) and Y$_2$O$_3$ (99.999\%) are used as the starting materials
for the host crystals. For doping, two types of europium oxide are employed: standard Eu$_2$O$_3$ with natural isotopic abundance for the 1\%-doped $\mathrm {{Eu}^{3+}}$:$\mathrm {Y_2SiO_5}$ crystal, and isotopically enriched $^{151}$Eu$_2$O$_3$ with 99.3\% $^{151}$Eu purity for the 700 ppm-doped $^{151}\mathrm {{Eu}^{3+}}$:$\mathrm {Y_2SiO_5}$ crystal.

To characterize the absorption spectrum, a bulk sample is cut from the 700 ppm-doped $^{151}\mathrm {{Eu}^{3+}}$:$\mathrm {Y_2SiO_5}$ crystal boule, with dimensions of 5.0 $\times$ 4.0 $\times$ 8.6 mm$^3$ along the $D_1 \times D_2 \times b$ axes. The absorption measurement is performed at 18 K, with the detection optical pulse polarized along the $D_1$ axis and propagating along the $b$ axis. The bulk crystal exhibits an optical inhomogeneous linewidth of 1.4 GHz and an absorption coefficient of 4.7 cm$^{-1}$.

\subsection*{2. The preparation of AFC with enhanced absorption}
According to Eq. 1 in the main text, enhancing the medium's absorption is crucial for improving the storage efficiency. However, creating AFCs with high absorption typically introduces strong slow-light dispersion, which in turn leads to an undesirably narrow storage bandwidth constrained by the slow-light cavity linewidth \cite{Sabooni_2013}. To solve this dilemma, we prepare AFCs with enhanced absorption embedded within a transparency window. As confirmed by theoretical simulations in Section 8, this approach enables the creation of AFCs with increased absorption by effectively mitigating slow-light effects.

\begin{figure}[tbph]
	\includegraphics [width=1\textwidth]{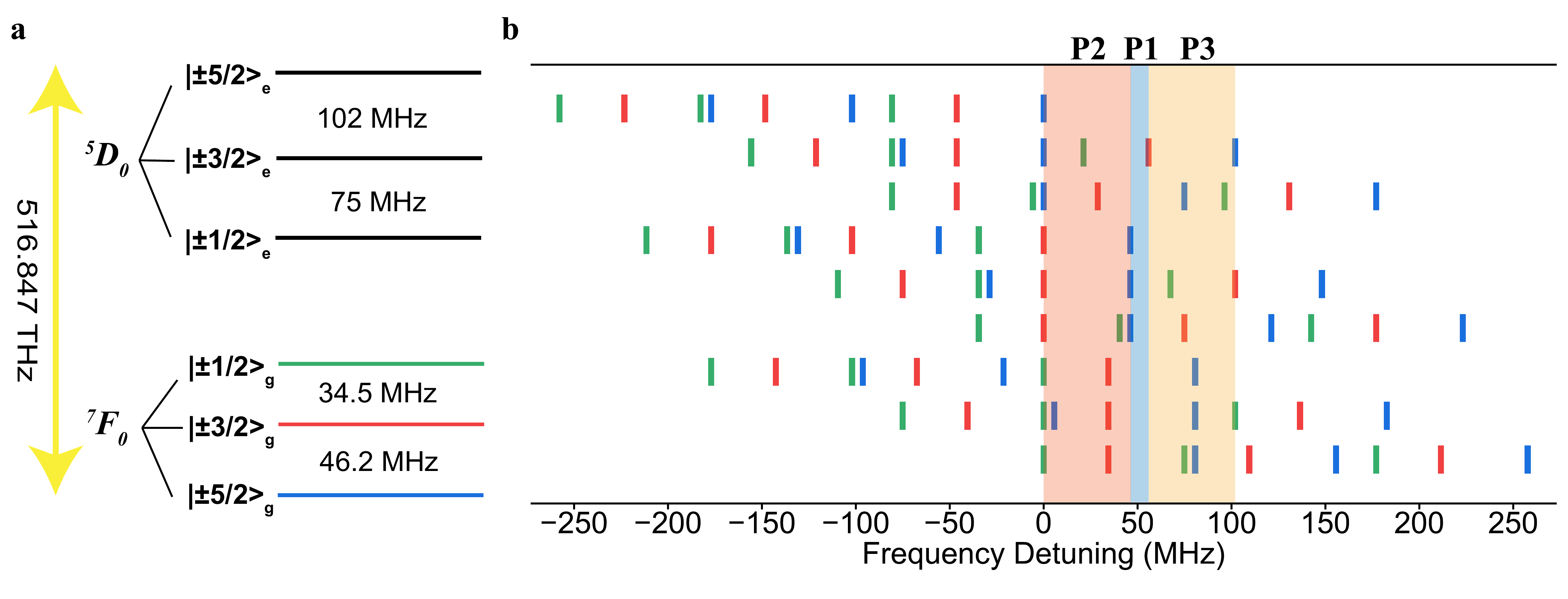}
	\caption{\textbf{a}, hyperfine structure of the $\mathrm{{^7}F{_0}\rightarrow{^5}D{_0}}$ transition for site-1 $\mathrm{^{151}Eu^{3+}}$ ions in $\mathrm{Y_2SiO_5}$ crystal at zero magnetic field. \textbf{b}, transition arrangement diagram for 9-classes site-1 $\mathrm{^{151}Eu^{3+}}$ ions in $\mathrm{Y_2SiO_5}$ crystal. Classification into these 9 classes depends on the transitions occurring when the incident light frequency is detuned to 0. From top to bottom, classes I to IX are shown. Green, red, and blue lines represent transitions from the ground states $|\pm1 /2\rangle_g$, $|\pm3/2\rangle_g$, and $|\pm5/2\rangle_g$, respectively. The three frequency pits are marked by shaded areas in orange, blue, and yellow, respectively. pit-2 (P2) and pit-3 (P3) are pumped in the preparation of the absorption-enhanced spectral hole-burning protocol, and pit-1 (P1) is the target band with enhanced absorption.}
	\label{fig:SM_energy_level}
\end{figure}

The detailed theoretical analysis of the pumping strategy is presented in Ref.~\cite{Zhu2022On-Demand}. 
Here, we analyze the absorption enhancement mechanism in this work. For site-1 $\mathrm{^{151}Eu^{3+}}$ ions in $\mathrm{Y_2SiO_5}$ crystal under zero magnetic field, the optical inhomogeneous broadening of $\mathrm{{^7}F{_0}\rightarrow{^5}D{_0}}$ transition significantly exceeds the hyperfine energy level spacing (Supplementary Fig.~\ref{fig:SM_energy_level}\textbf{a}), rendering the hyperfine transitions indistinguishable in the optical spectrum. 
An input laser resonating with the $\mathrm{{^7}F{_0}\rightarrow{^5}D{_0}}$ transition will excite 9 distinct transitions connecting all hyperfine states, which correspond to 9 classes of $\mathrm{^{151}Eu^{3+}}$ ions. 
Utilizing the transition arrangement diagram for 9 classes of $\mathrm{^{151}Eu^{3+}}$ ions (Supplementary Fig.~\ref{fig:SM_energy_level}\textbf{b}), we can identify spectral hole and anti-hole frequencies under zero-detuning optical excitation.
This provides the foundation for designing the absorption-enhancement protocol. 
Taking Class I ions as an example (the top one in Supplementary Fig.~\ref{fig:SM_energy_level}\textbf{b}), incident light resonantly drives the $|\pm5/2\rangle_g\rightarrow|\pm5/2\rangle_e$ transition, transferring population from the $|\pm5/2\rangle_g$ state to the other two ground-state energy levels. Consequently, the frequency of the other two blue-marked transitions exhibits spectral holes, while the frequency of other color-coded transitions generates anti-holes.
When considering a spectrally broadened pit, all transitions in Supplementary Fig.~\ref{fig:SM_energy_level}\textbf{b} can be represented with identical linewidths. Throughout the analysis, we conventionally adopt rightward spectral broadening (toward higher frequencies) as the default configuration.

As illustrated in Supplementary Fig.~\ref{fig:AFC sequences}, the ``anti-hole burning" process is performed following the initialization process. Two chirped pulses, with frequency ranges of [0, $46.2$ MHz] (denoted as P2) and [$55.8$ MHz, $102$ MHz] (denoted as P3), are applied to generate an antihole within the frequency range of [$46.2$ MHz, $55.8$ MHz] (denoted as P1). All transitions are considered to undergo $102$ MHz rightward broadening. Here, we explain the analysis procedures, taking class IV ions as an example.

\begin{figure}[tbph]
	\includegraphics [width=0.6\textwidth]{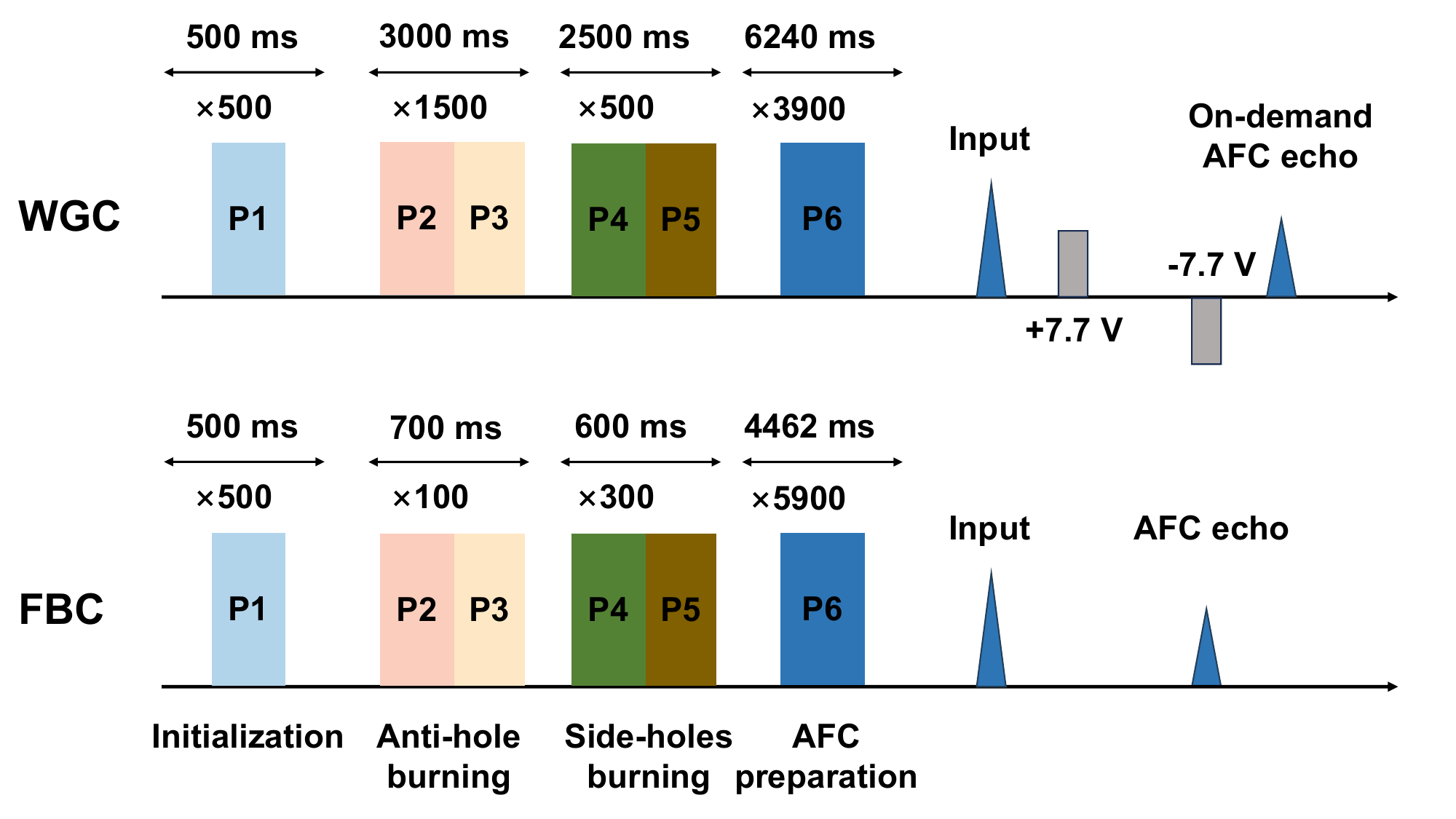}
	\caption{Time sequences for AFC quantum memories in the WGC and FBC schemes. The different steps are assigned distinct colors and implemented in dedicated frequency pits (labeled P1 through P6), with corresponding frequency ranges (relative to zero detuning) as follows: P1 is [46.2 MHz, 55.8 MHz], P2 is [0 MHz, 46.2 MHz], P3 is [55.8 MHz, 102 MHz], P4 is [44 MHz, 48 MHz], P5 is [54 MHz, 58 MHz], P6 is [48 MHz, 54 MHz]. P1, P2, P3 are also indicated with the same color in the transition diagram shown in Supplementary Figure~\ref{fig:SM_energy_level}. We create isolated antiholes with enhanced absorption while suppressing absorption in the surrounding regions, enabling both efficient and wideband cavity-enhanced AFC storage simultaneously. After the AFC preparation process, two additional electric pulses are applied in the WGC scheme to facilitate on-demand retrieval via Stark-shift-induced interference.}
	\label{fig:AFC sequences}
\end{figure}

\begin{itemize}
\item \textbf{Step 1: Analysis of the pumped frequency ranges.} The $|\pm3/2\rangle_g\rightarrow|\pm5/2\rangle_e$ transition is pumped within the frequency ranges [0, 46.2 MHz] and [55.8 MHz, 102 MHz]; The $|\pm5/2\rangle_g\rightarrow|\pm5/2\rangle_e$ transition is pumped within the frequency ranges [9.6 MHz, 55.8 MHz]; The $|\pm1/2\rangle_g\rightarrow|\pm5/2\rangle_e$ transition is pumped within the frequency ranges [34.5 MHz, 80.7 MHz] and [90.3 MHz, 102 MHz]; The $|\pm3/2\rangle_g\rightarrow|\pm3/2\rangle_e$ transition is pumped within the frequency ranges [55.8 MHz, 102 MHz].
\item \textbf{Step 2: Analysis of the frequency range of transitions that contribute to the absorption in P1.} The $|\pm5/2\rangle_g\rightarrow|\pm5/2\rangle_e$ transition within the frequency range [0, 9.6 MHz]; The $|\pm3/2\rangle_g\rightarrow|\pm5/2\rangle_e$ transition within the frequency range [46.2 MHz, 55.8 MHz]; The $|\pm1/2\rangle_g\rightarrow|\pm5/2\rangle_e$ transition within the frequency range [80.7 MHz, 90.3 MHz].
\item \textbf{Step 3: Based on the hole-burning results from Step 1, determining the contributions of the transitions in Step 2.}
In the frequency range [0, 9.6 MHz], the population is distributed in $|\pm1/2\rangle_g$ and $|\pm5/2\rangle_g$, so the $|\pm5/2\rangle_g\rightarrow|\pm5/2\rangle_e$ transition contributes to uniform absorption for P1;
In the frequency range [46.2, 55.8 MHz], the population is distributed in $|\pm3/2\rangle_g$, so the $|\pm3/2\rangle_g\rightarrow|\pm5/2\rangle_e$ transition contributes to uniform absorption for P1;
In the frequency range [80.7, 90.3 MHz], the population is distributed between $|\pm1/2\rangle_g$ and $ |\pm 5/2\rangle_e$, so the $|\pm1/2\rangle_g\rightarrow|\pm5/2\rangle_e$ transition contributes to uniform absorption for P1.
By sequentially analyzing all classes of ions, we observe an absorption enhancement within the [46.2 MHz, 55.8 MHz] frequency range.
To avoid double-counting the absorption contributions from different classes of ions, we reassign these contributions to different transitions, as summarized in Supplementary Tab.~\ref{Tab.Trans_Contribution}. Taking into account the branching ratios for transitions (shown in Supplementary Tab.~\ref{Tab.BR}), we quantify the magnitude of absorption enhancement relative to the natural absorption. The absorption is estimated to be enhanced 2.87 times in the frequency range [46.2 MHz, 50.1 MHz], 2.97 times in the frequency range [50.1 MHz, 51.9 MHz], and 2.55 times in the frequency range [51.9 MHz, 55.8 MHz].

\end{itemize}  

In the WGC (and similarly in the FBC scheme), each chirped pulse has a duration of 1 ms (3.5 ms for the FBC) and is repeated 1500 times (100 times for the FBC). Following this process, a $\sim$2.7-fold increase in the absorption depth, $d$, is observed in the WGC, with $d$ increasing from $\sim$0.97 to $\sim$2.6. In the FBC, it is inferred that $d$ would increase from $\sim$0.052 to $\sim$0.14, although it cannot be directly measured.

\begin{table}[tbph]
    \caption{\label{Tab.Trans_Contribution} The estimated contribution to the absorption in the frequency range [46.2 MHz, 55.8 MHz] from individual optical transitions. We assume ions are initially uniformly distributed across the three ground states. If one ground state is burned, we simply assume the population of the remaining two ground states increases to 1.5 times of their original values. Similarly, if two ground states are hole burning, the population of the remaining ground state increases to 3 times its initial value.}
    \begin{ruledtabular}
    \begin{tabular}{ccc}
        Transitions & Estimated enhancement for the frequency ranges\\
        \hline
        $|\pm1/2\rangle_g\rightarrow|\pm1/2\rangle_e$ & 1.5 times for [46.2 MHz, 55.8 MHz]\\
        $|\pm1/2\rangle_g\rightarrow|\pm3/2\rangle_e$ & 1.5 times for [46.2 MHz, 50.1 MHz], and 3 times for [50.1 MHz, 55.8 MHz]\\
        $|\pm1/2\rangle_g\rightarrow|\pm5/2\rangle_e$ & 3 times for [46.2 MHz, 55.8 MHz]\\
        $|\pm3/2\rangle_g\rightarrow|\pm1/2\rangle_e$ & 3 times for [46.2 MHz, 55.8 MHz]\\
        $|\pm3/2\rangle_g\rightarrow|\pm3/2\rangle_e$ & 3 times for [46.2 MHz, 55.8 MHz]\\
        $|\pm3/2\rangle_g\rightarrow|\pm5/2\rangle_e$ & 3 times for [46.2 MHz, 55.8 MHz]\\
        $|\pm5/2\rangle_g\rightarrow|\pm1/2\rangle_e$ & 3 times for [46.2 MHz, 51.9 MHz], and 1.5 times for [51.9 MHz, 55.8 MHz]\\
        $|\pm5/2\rangle_g\rightarrow|\pm3/2\rangle_e$ & 3 times for [46.2 MHz, 55.8 MHz]\\
        $|\pm5/2\rangle_g\rightarrow|\pm5/2\rangle_e$ & 1.5 times for [46.2 MHz, 55.8 MHz]\\
    \end{tabular}
    \end{ruledtabular}
\end{table}

\begin{table}[tbph]
    \caption{\label{Tab.BR} The branching ratio of the $\mathrm{{^7}F{_0}\rightarrow{^5}D{_0}}$ transition for site-1 $\mathrm{^{151}Eu^{3+}}$ ions in $\mathrm{Y_2SiO_5}$ crystal at zero magnetic field. The measurement was performed using spectral hole burning, similar to that presented in Ref.~\cite{Lauritzen2012Spectroscopic}.}
    \begin{ruledtabular}
    \begin{tabular}{cccc}
        ~ & 
        $|\pm1/2\rangle_e$ & 
        $|\pm3/2\rangle_e$ & 
        $|\pm5/2\rangle_e$ \\
        \hline
        $|\pm1/2\rangle_g$ & 
        $0.03$ &
        $0.21$ & 
        $0.76$ \\
        $|\pm3/2\rangle_g$ & 
        $0.12$ &
        $0.67$ & 
        $0.21$ \\
        $|\pm5/2\rangle_g$ & 
        $0.85$ &
        $0.12$ & 
        $0.03$ \\
    \end{tabular}
    \end{ruledtabular}
\end{table}


Next, the ``side-holes burning" process is implemented to remove the absorption around the center antihole. Two chirped pulses, with frequencies ranges of [44 MHz, 48 MHz] (denoted as P4) and [54 MHz, 58 MHz] (denoted as P5), are applied to create an isolated antihole within the frequency range of [48 MHz, 54 MHz] (denoted as P6). In the WGC (and similarly in the FBC), each chirped pulse has a duration of 2.5 ms (1 ms for the FBC) and is repeated 500 times (300 times for the FBC).

Subsequently, a parallel comb preparation scheme \cite{Jobez2016Towards} is employed to construct an AFC structure with a comb spacing $\Delta$ of 1 MHz in the WGC and 0.5 MHz in FBC, respectively, within the 6 MHz bandwidth of the isolated antihole. The parameters of the parallel comb preparation scheme are optimized using a Python program to maximize the storage efficiency, where impedance-matched conditions should be fulfilled. In the WGC (and similarly in the FBC), each AFC preparation pulse lasts for 1.6 ms (1 ms for the FBC) and is repeated 3900 times (4462 for the FBC). 
For the FBC-based quantum memory, the cavity's resonance is highly sensitive to vibrations caused by the cryostat. To mitigate this, both the AFC preparation and signal pulse inputs are synchronized within low-vibration windows, which account for approximately 70\% of each 997-ms vibration cycle. While the comb structures inside the cavity cannot be directly observed, the AFC finesse is estimated to be 12 for the WGC and 8 for the FBC, based on the impedance-matched condition.

In the WGC, we further demonstrate on-demand retrieval of the AFC memory via active control of Stark-shift-induced interference between sub-ensembles of $^{151}\mathrm{Eu}^{3+}$ ions \cite{Horvath2021Noise-free,Liu2020On-demand}. The Stark coefficient of $^{151}\mathrm{Eu}^{3+}$ was previously determined in Ref. \cite{Liu2020On-demand}. Two electric pulses with opposite polarities are applied to electrodes positioned on either side of the optical waveguide. The first electric pulse is applied after the signal pulse at the time $t=0$ and before $t=1/\Delta$, suppressing the standard first AFC echo by inducing destructive interference between two sub-ensembles of $^{151}\mathrm{Eu}^{3+}$ ions with opposite Stark responses. The second electric pulse is applied within the time window ($(n-1)/\Delta$, $n/\Delta$), canceling the relative phase and enabling the on-demand retrieval of the $n$-th echo at $t=n/\Delta$. The electric pulse has a full width at half maximum (FWHM) of 48 ns and a voltage of 7.7 V, inducing a relative phase shift of $\pi$ between the two sub-ensembles of $^{151}\mathrm{Eu}^{3+}$ ions.

\subsection*{3. The characterization and fabrication of ultra-low-loss optical waveguides}
To achieve efficient quantum storage in the WGC, minimizing the propagation loss of the optical waveguide is essential. Previously, we identified a set of parameters for the FLM process that successfully reduced these losses \cite{Su2022On-demand}. However, further optimization poses significant challenges due to the difficulty in accurately measuring propagation losses when they are extremely low. Additionally, it is difficult to distinguish propagation losses from coupling losses solely through transmission measurements of the optical waveguides.

\begin{figure}[tbph]
	\includegraphics [width=0.6\textwidth]{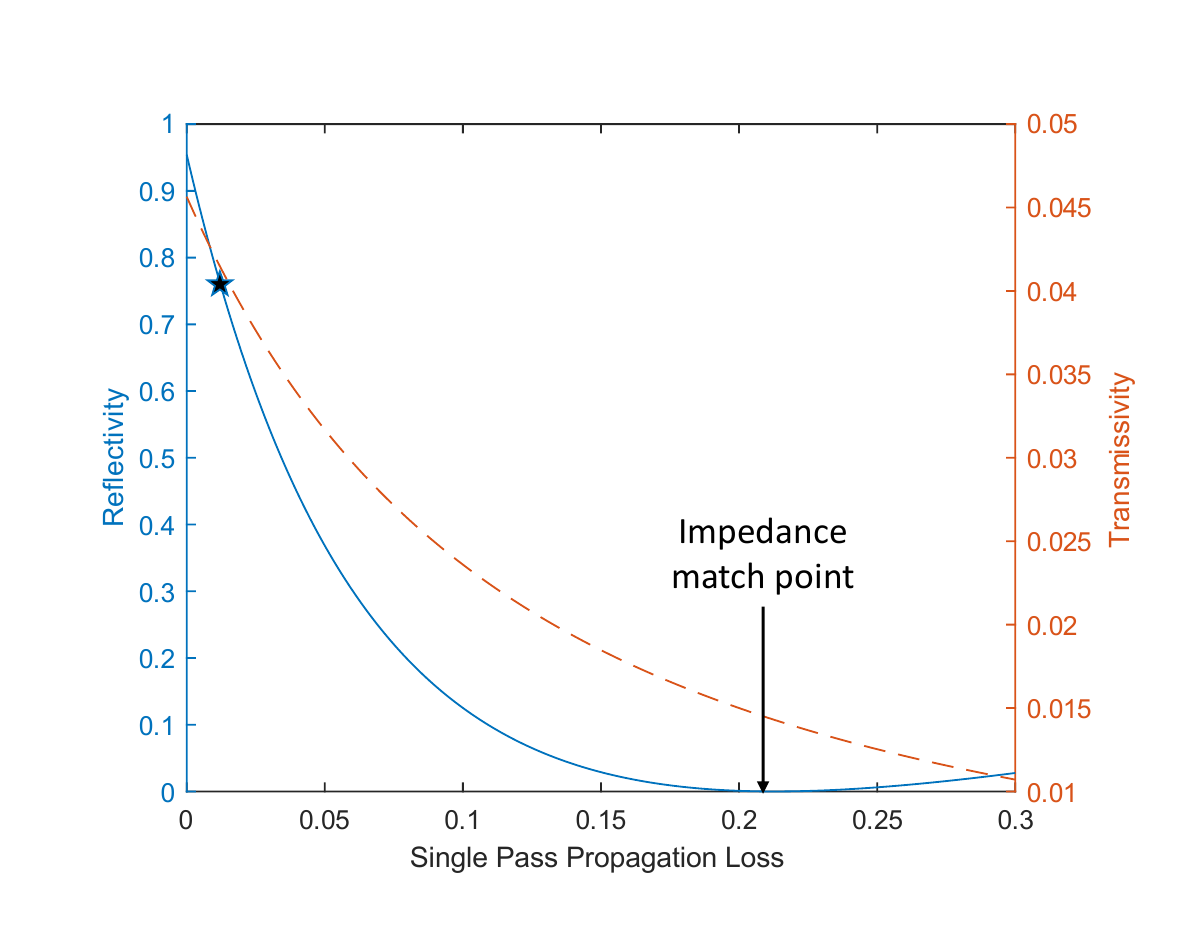}
	\caption{The reflectivity and transmissivity of a waveguide-based cavity as functions of propagation loss, calculated for mirror reflectivities of $R_1$=0.65, $R_2$=0.995. The black star indicates the measured loss of the optimized waveguide, which is 0.012 over a length of 4.9 mm.}
	\label{fig:wgloss reducing}
\end{figure}

To address this challenge, we develop a novel method to accurately measure propagation losses by utilizing the cavity effect. The reflectivity $\mathcal{R}$ and transmissivity $\mathcal{T}$ of a Fabry-Perot cavity, accounting for a finite single-pass propagation loss $d_{\mathrm{loss}}$, can be expressed under resonance conditions as follows: 
\begin{equation}
\begin{aligned}
\mathcal{R}&=1-\frac{(1-R_1)(1-R_2 e^{-2(d_{\mathrm{loss}}+\tilde{d})})}{(1-\sqrt{R_1 R_2} e^{-(d_{\mathrm{loss}}+\tilde{d})})^2} \cdot \eta_{\mathrm{M}}\\
\mathcal{T}&=\frac{(1-R_1)(1-R_2)}{(1-\sqrt{R_1R_2}e^{-(d_{\mathrm{loss}}+\tilde{d})})^2} \cdot \eta_{\mathrm{M}},
\end{aligned}
\label{RandT}
\end{equation} 
as illustrated in Supplementary Figure~\ref{fig:wgloss reducing} with $\eta_{\mathrm{M}}=1$ and $\tilde{d}=0$. 
Note that here the round-trip intra-cavity loss is primarily attributed to propagation losses and the transmissivity of the second mirror $T_2=1-R_2$, with the approximation of $\epsilon \approx T_2+2d_{\mathrm{loss}}$.
When the propagation loss is significantly lower than the impedance matching point, the cavity’s reflectivity becomes highly sensitive to changes in propagation loss. 

By measuring the resonance reflectivity of the WGC outside the absorption profile, one can precisely evaluate the performance of the waveguide. Additionally, under the impedance-matched condition $R_1=R_2 e^{-2(d_{\mathrm{loss}}+\tilde{d})}$, we measure a $\eta_{\mathrm{M}}=98.5\%$ as determined from the resonance reflectivity $\mathcal{R}=1-\eta_{\mathrm{M}}$. The imperfection in $\eta_{\mathrm{M}}$ is due to slight non-Gaussian profile of the waveguide mode. Using this method, we carefully optimize the parameters of the FLM process on coated crystals, successfully fabricating ultra-low loss waveguides with a single-pass propagation loss $d_{\mathrm{loss}}$ of 0.012 (equivalent to 0.052 dB, see Supplementary Figure~\ref{fig:wgloss reducing}). This optimization results in an intra-cavity loss $\epsilon=0.029$. Given the loss of both mirrors, the finesse of WGC in the absence of absorption loss can be calculated via $\mathcal{F} = \frac{\pi \sqrt[4]{R_1 \epsilon}}{1 - \sqrt{R_1 \epsilon}}=13.6$. With the crystal length of 4.9 mm, the WGC exhibits a free spectral range of 17 GHz and a linewidth of 1.25 GHz. 

Reaching 90\% storage efficiency demands $\epsilon$ down to 0.01. A practical route is to lower the transmissivity $T_{2}$ of the second mirror with higher-reflectivity coatings. The single-pass propagation loss $d_{\mathrm{loss}}$ can be reduced by further optimizing fabrication parameters, adopting alternative waveguide designs (e.g., type-III waveguides), or performing post-fabrication annealing to relieve inhomogeneous broadening and simultaneously increase absorption and decrease loss.

As shown in Supplementary Figure~\ref{fig:WGC setup}b and \ref{fig:WGC setup}c, Type IV waveguides are fabricated on the upper surface of the $^{151}\mathrm {{Eu}^{3+}}$:$\mathrm {Y_2SiO_5}$ crystal. Each waveguide consists of three coplanar notches on both sides, with the middle notch positioned 19 \textmu m away from the crystal surface. Both the height and width of the waveguides are 23.3 \textmu m. Femtosecond laser pulses with a wavelength of 1030 nm, a pulse duration of 210 fs, and a repetition rate of 201.9 kHz are employed for fabrication, focused through a 50× objective lens (Numerical aperture, NA = 0.65). The pulse energies used are 71.6 nJ, 69.8 nJ, and 68.4 nJ for the lower, middle, and upper notches, respectively. The waveguide is etched along the b axis of the $\mathrm {Y_2SiO_5}$ crystal on the $D_1-b$ surface, with the etching pulses polarized parallel to the b axis. Two electrodes are fabricated through ultraviolet lithography (Karl Suss, MABA6) and electron beam evaporation (K.J. Lesker, LAB 18).

The peak absorption depth of our bulk crystal is initially 2.3, accompanied by an inhomogeneous broadening of 1.4 GHz. After the fabrication of the waveguide, the absorption depth decreases to typical values of 1.3, primarily due to additional broadening induced by internal stresses from the fabrication process. In the final experiments, the absorption depth further decreases slightly to approximately 0.97 due to broadening induced by the external pressure exerted by the thermal conductive strips.

\subsection*{4. Experimental setup for the WGC-based quantum memory}

\begin{figure}[tbph]
	\includegraphics [width=1.0\textwidth]{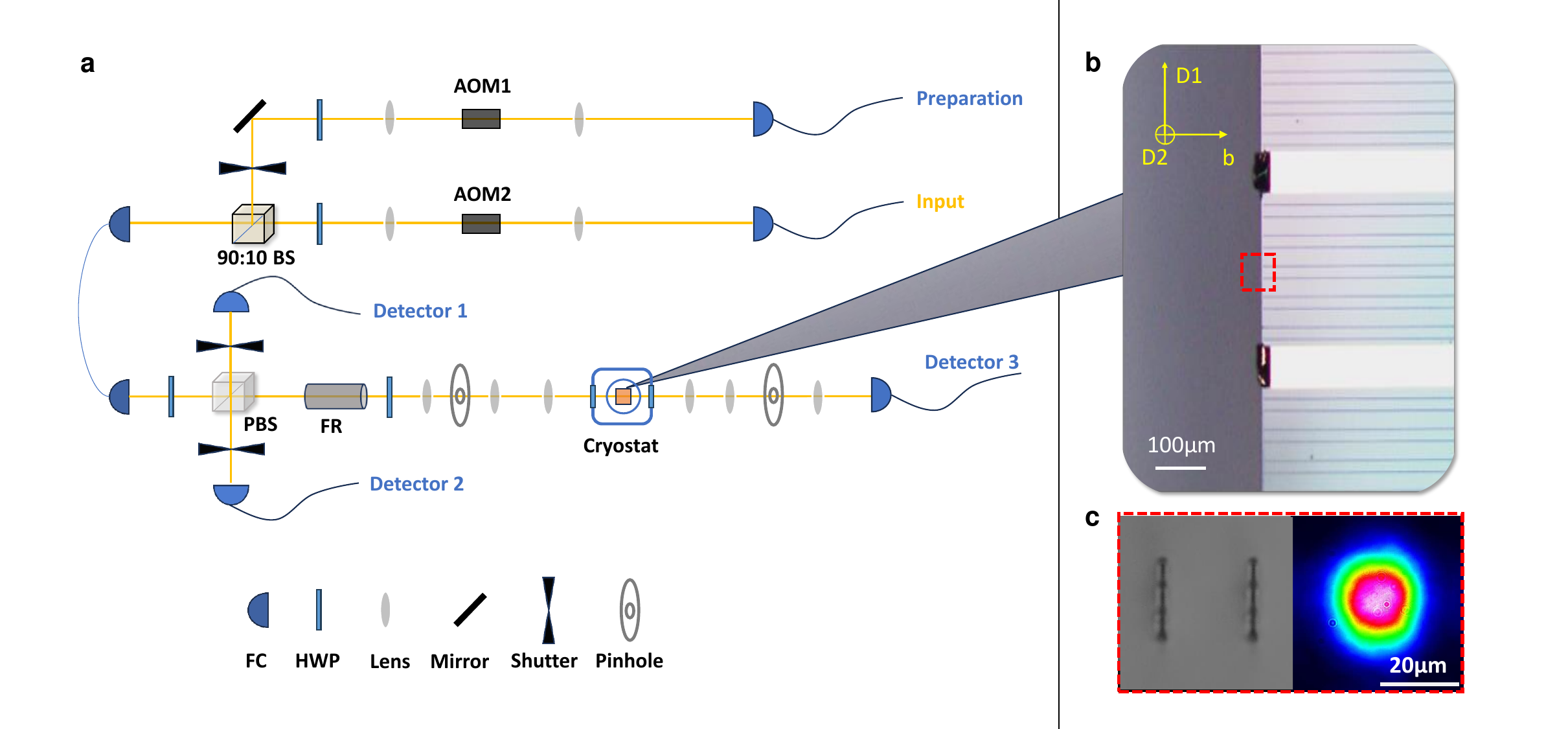}
	\caption{\textbf{a.} Experimental setup for the WGC. The input and preparation beams are generated using acousto-optic modulators (AOMs) in a double-pass configuration, controlled by an arbitrary waveform generator (HDAWG, Zurich Instruments). These beams are combined using a beam splitter (BS) and then coupled into a single-mode fiber. A silicon-based single-photon detector (Si-SPD, Detector 1) monitors the input signal strength as a reference. The input signal, polarized along the crystal's D1 axis, is coupled into the waveguide using a lens group. A polarization beam splitter (PBS) and a Faraday rotator (FR) separate the input signal and the retrieved echoes based on their respective polarization states. Both the retrieved echoes and reflected light are detected by a second Si-SPD (Detector 2). Shutters are employed to protect the Si-SPDs during the preparation procedures, while pinholes spatially reject spurious light. FC: fiber coupler. HWP: half-wave plate. \textbf{b}. Top view of the device under a microscope. Five type-IV optical waveguides are fabricated between two electrodes, with the central waveguide used in the quantum memory experiments. A 100 \textmu m scale bar is included for reference. \textbf{c}. Front view of the type-IV waveguides and the beam profile of the guided mode at the rear facet of the waveguide cavity. A 20 \textmu m scale bar is included for reference.}
	\label{fig:WGC setup}
\end{figure}

The experimental setup for the WGC-based quantum memory is illustrated in Supplementary Figure~\ref{fig:WGC setup}a. The waveguide cavity is cooled to approximately 3.6 K by a closed-cycle cryostat (Montana Instruments). The input signal and the backward retrieved echo are separated by a polarization beam splitter (PBS) with a Faraday rotator (FR). The retrieved echo is detected by a single-photon detector (Detector 2), while a reference single-photon detector (Detector 1) records the reference signal for the input. Transmitted light from the waveguide cavity is detected by a photodetector (Detector 3).

\begin{figure}[tbph]
	\includegraphics [width=0.6\textwidth]{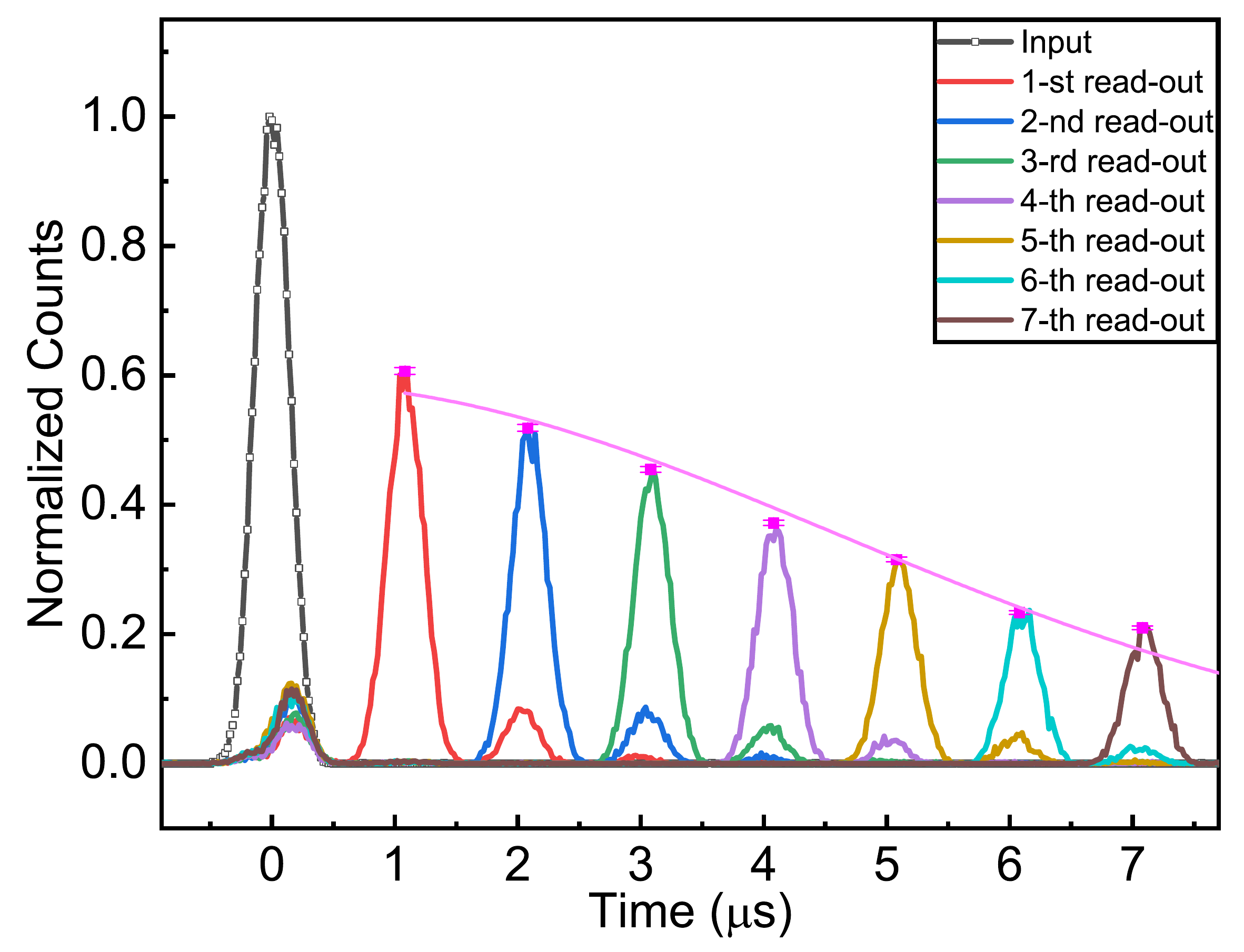}
	\caption{Optimizing the storage efficiency for high-order AFC echoes in the WGC with $\mu = 0.35$. The highest storage efficiency for the 7th echo is measured to be 21.0(3)\%. The pink solid line represents the fitted efficiency decay as described by Eq.~\ref{SMAFC efficiency} in Sec.~4 of the Supplementary Information, with a fitted finesse $F_{\mathrm{AFC}}$ of 17.0(2).}
	\label{fig:SM_WGC_fig4_7-thmax}

\end{figure}

To facilitate the measurement of input signals and the absorption depth of our sample, a portion of the rear facet (the second mirror) is left uncoated. When the input beam is coupled into the waveguide at this blocked cavity region, 65\% of the light is reflected by the first mirror ($R_1=65\%$) and detected by Detector 2, while the transmitted portion is absorbed by the sample and subsequently measured by Detector 3. The absorption depth can be determined by measuring the sample transmission variation inside and outside the absorption profile. In such case, the actual input corresponds to the signal detected by Detector 2 divided by $R_1$. We adjust the half-wave plate (HWP) to ensure the reference signal measured by Detector 1 matches this value, and treat this reference signal as the input for our analysis.

The WGC is inherently vibration-robust and requires no cavity-length locking. For simplicity we therefore made no vibration-isolation effort in the WGC experiment, resulting in a comparatively short optical-coherence lifetime of 160 \textmu s versus that of the FBC.

The efficiency of cavity-enhanced Stark-modulated AFC echoes, $\eta_{\mathrm{cav}}$, is described by \cite{Horvath2021Noise-free}:
\begin{equation}
\eta_{\mathrm{cav}}=\frac{4\tilde{d}^2 e^{-2 \tilde{d} }\left(1-R_1\right)^2 R_2}{\left(1-\sqrt{R_1 R_2} e^{-\tilde{d}}\right)^4} (\eta_{\mathrm{deph}})^{n^2},
\label{SMAFC efficiency}
\end{equation}
where $n$ is the order of the retrieved echo, $\eta_{\mathrm{deph}}=e^{-\frac{\pi^2/2\mathrm{ln}2}{F^2_{\mathrm{AFC}}}}$ accounts for the dephasing due to finite combs, assuming a Gaussian lineshape. Using Eq.~\ref{SMAFC efficiency}, we determine a fitted finesse of 11.9(1) in Fig.~2b of the main text, which matches the estimated value of 12 derived from the impedance-matched condition in the WGC. The efficiency of AFC echoes is highly sensitive to $F_{\mathrm{AFC}}$, and efficient higher-order AFC storage can be achieved with increased $F_{\mathrm{AFC}}$. With a slight deviation from the impedance-matched condition, we prepare AFC with a finesse of $17.0(2)$, and achieve an efficiency improvement from 7.4(2)\% to 21.0(3)\% for the 7th echo, while maintaining an efficiency of 60.6(6)\% for the 1st echo (Supplementary Figure~\ref{fig:SM_WGC_fig4_7-thmax}).

\subsection*{5. The fabrication and stabilization of FBC}\label{sec:match}

The fiber-based microcavity consists of a fiber concave mirror and a membrane planar mirror, both of which are coated with stacks of alternating high-index
(Ta$_2$O$_5$) and low-index (SiO$_2$) thin films to form distributed Bragg reflectors.
Using CO$_2$ laser machining technique \cite{Hunger_2010}, a fiber concave mirror with a roughness of approximately 0.2 nm is fabricated. This process involves focusing multiple ablation pulses onto the surface of a multimode fiber, creating a nearly Gaussian-shaped depression. The resulting concave surface is subsequently coated, exhibiting a radius of curvature (ROC) of approximately 820 \textmu m and an effective mirror diameter of about 35 \textmu m. A typical surface of the fiber concave mirror is shown in Supplementary Figure~\ref{fig:fiber cavity}a. For the membrane planar mirror, we start by coating one side of the $D_{1} \times D_{2}$ planes of a $^{151}\mathrm {{Eu}^{3+}}$:$\mathrm {Y_2SiO_5}$ bulk crystal with  reflective films. Subsequently, the opposite side of the bulk crystal is polished to reduce its thickness along the $b$-axis to approximately $L_{\rm memb}=200$ \textmu m (or 70 \textmu m), using a target surfacing system (Leica EM TXP). The polished surface, parallel to the coated surface, demonstrates a roughness of approximately 0.3 nm, as measured by atomic force microscopes over a $10 \times 10$ \textmu m$^2$ detection area. As shown in Supplementary Figure~\ref{fig:fiber cavity}b, the assembly of the FBC involves bonding the fiber tail of the fiber concave mirror and the coated surface of the membrane planar mirror onto a shear piezoelectric ceramic stack and a sapphire substrate with a central hole, respectively. Both components are subsequently fixed onto a shared titanium alloy base to align the two mirrors. The air gap between the fiber concave mirror and the membrane measures approximately 100 \textmu m. Given the total loss of approximately 3.67\% in the absence of membrane absorption, the finesse of bare FBC can be calculated to be 171. For an intra-cavity optical path length of $200 \times 1.8 + 100 = 460~\text{\textmu m}$, the FBC exhibits a free spectral range of 326 GHz and a linewidth of 1.9 GHz.


\begin{figure}[tbph]
\includegraphics [width=0.8\textwidth]{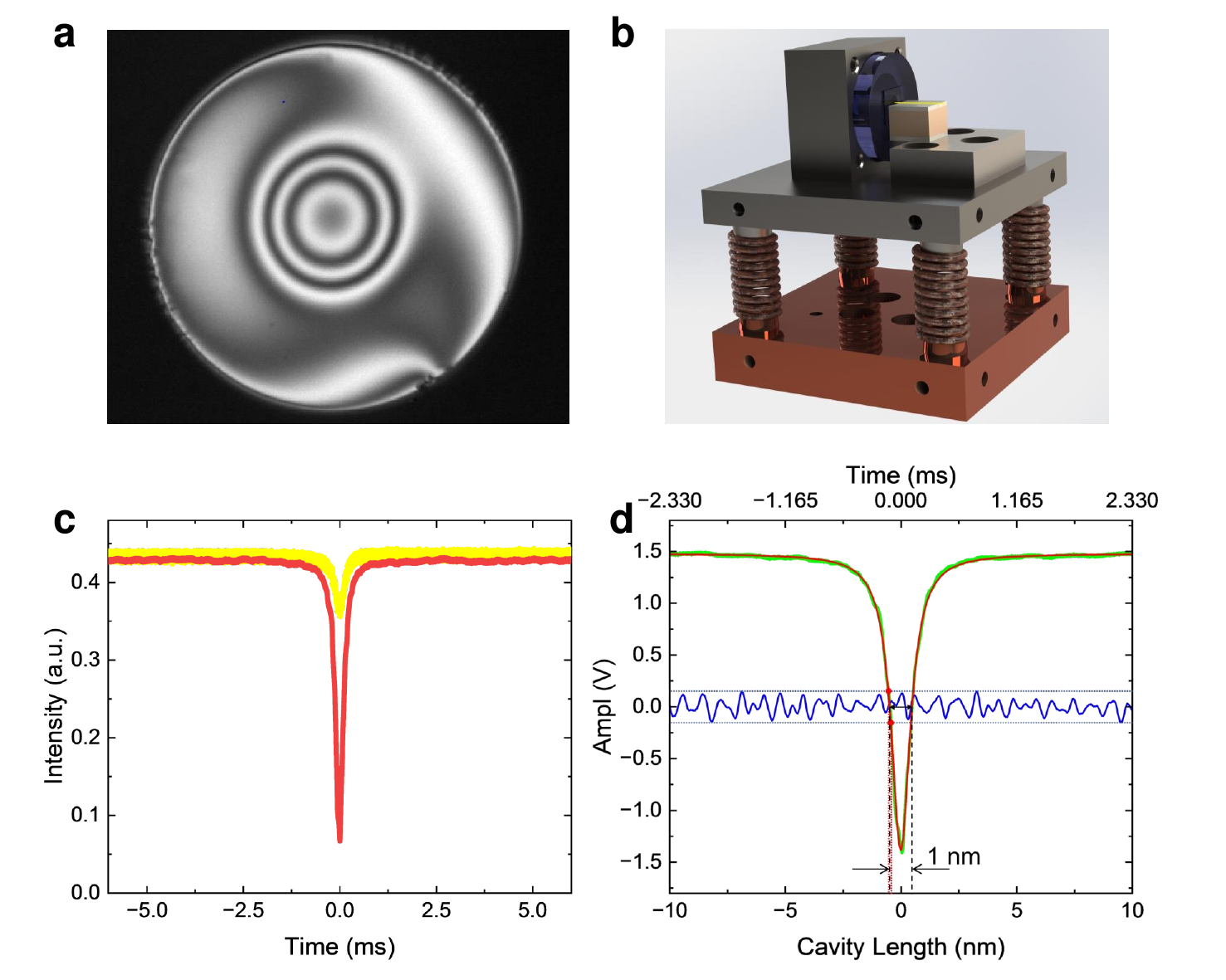}
\caption{\textbf{a}. A typical surface of a fiber concave mirror. \textbf{b}. The structure of the vibration-isolation system. The base beneath the springs is made of oxygen-free copper for efficient heat conduction, while the base above the springs is made of a titanium alloy to retain high hardness at cryogenic temperatures. \textbf{c}. The intensity variation of reflection signals from the reference (red) and stored (yellow) light with the cavity length scanned over time by a shear piezoelectric actuator. \textbf{d}. A typical performance of a stabilized FBC operating at 4.3 K, achieved with a stronger connection of copper strips. The amplitude of the error signal is plotted against the scanned (green) and locked (blue) cavity length over time (top axis). The resonance dip in the error signal is modeled using the Lorentzian function (red), with the time axis converted to the cavity length axis (bottom axis) based on the resonance linewidth of approximately 1 nm for the 606-nm reference light. The amplitude jitter in the error signal here amounts to approximately 10\% of the peak-to-peak value of the resonance dip. Based on the fitted Lorentzian function, an inverse transformation from voltage to cavity length is implemented, yielding an equivalent cavity length jitter of 100 pm for the data presented.}
\label{fig:fiber cavity}
\end{figure}

To optimize the mode-matching efficiency $\eta_{\mathrm{M}}$, the input light is focused onto the membrane planar mirror and coupled into the cavity. For the  Gaussian cavity mode (TEM$_{00}$), $\eta_{\mathrm{M}}$ is given by: 
\begin{equation}
\begin{aligned}
\label{eq:mode}
\eta_{\mathrm{M}}\approx(\frac{2w_0w_c}{w_0^2+w_c^2})^2,
\end{aligned}
\end{equation}
where $w_0$ is the waist of the Gaussian cavity mode, and $w_c$ is the waist of the input Gaussian beam. Achieving $\eta_{\mathrm{M}}\approx 1$ requires $w_0 \approx w_c$. For the cavity geometry, $w_0$ is calculated as $w_0=\sqrt{\lambda/\pi} \cdot (L^{\prime}({\rm ROC}-L^{\prime}))^{\frac{1}{4}} \approx 8.14$ \textmu m. Here, $L^{\prime}$ represents the effective cavity length, accounting for the refraction of intra-cavity light within the membrane. It is calculated as $L^{\prime}={L_{\rm memb}}/{n_{\rm memb}}+L_{\rm air} \approx 211.1$ \textmu m, where $n_{\rm memb}$ denotes the refractive index of the membrane \cite{vandam2018}. Using a two-fold beam expander and a 100-mm focal length lens (see Supplementary Figure~\ref{fig:FIG1}), the input beam waist is adjusted to $w_c \approx 7.95$ \textmu m, yielding a theoretical mode-matching efficiency of $\eta_{\mathrm{M}}=99.9\%$. Similar to that in the WGC, we measure $\eta_{\mathrm{M}}$ through the resonance reflectivity under the impedance-matched condition. To search for the impedance-matched point, we simultaneously scan the cavity length and the input light frequency across the membrane's absorption profile at 580 nm. This yields a minimum reflectivity $\mathcal{R}$ of 5\%, which corresponds to a measured $\eta_{\rm M}$ of 95\%. The small deviation from the theoretical value is attributed to aberrations induced by the lens, misalignments between the microcavity inside the cryostat and the optical path outside the cryostat, and imperfections in the perpendicular alignment between the fiber end and the membrane.

A low ratio of $\epsilon/{4\widetilde{d}}$ is the prerequisite for achieving high-efficiency quantum memory, according to Eq. 1 in the main text. The intra-cavity loss $\epsilon$ primarily comprises scattering loss at the membrane-air interface, clipping loss at the fiber concave mirror, and transmission loss through the fiber concave mirror. 
Based on the surface roughness measurements, the scattering loss at the membrane-air interface is estimated to be approximately 40 ppm \cite{10.1063/1.3679721}. Clipping loss, due to the limited effective mirror diameter of the fiber concave mirror, is estimated to be 1035 ppm. The transmission loss through the fiber concave mirror at 580 nm is designed to be below 500 ppm. According to Eq.~(\ref{RandT}), the actual intra-cavity loss $\epsilon$ can be deduced from the resonance reflectivity measured at 580 nm, outside the membrane's absorption profile (i.e., $\widetilde{d}=0$). The measured resonance reflectivity $\mathcal{R}$ is 84.1\%, with a mode-matching efficiency $\eta_{\rm M}$ of 95\%. Consequently, $\epsilon$ is calculated as 1680 ppm. The small deviation compared to the theoretical value can be attributed to the imperfection of the coating. 

Achieving high-efficiency quantum storage in the FBC, which is an open cavity, requires minimizing vibrations and stabilizing cavity length. To fulfill this requirement, we employ a combination of passive vibration isolation and active cavity length stabilization techniques. Specifically, four springs serve as passive isolators to mitigate high-frequency cryostat vibrations, as illustrated in Supplementary Figure~\ref{fig:fiber cavity}b. Flexible copper strips (not shown in Supplementary Figure~\ref{fig:fiber cavity}b), serving as heat conductors, connect the upper and lower bases of springs to ensure effective membrane cooling. The membrane, bonded on a sapphire substrate being cooled to 4.7 K, demonstrates an optical coherence lifetime of approximately 600 \textmu s measured by the two-pulse photon echo. Additionally, both the fiber concave mirror and the membrane are securely mounted on a shared titanium alloy base to efficiently mitigate low-frequency vibrations. For active stabilization of the cavity length, the fiber concave mirror is attached to a shear piezoelectric ceramic stack, which provides a maximum stroke of approximately 4 \textmu m at room temperature. Due to the poor thermal conductivity of the titanium alloy base located above the springs, the temperature of the shear piezoelectric ceramic stack is higher than that of the sample (4.7 K), which is directly connected to flexible copper strips. As a result, the stroke of the shear piezoelectric ceramic stack remains relatively sufficient at cryogenic temperatures, around 150 nm. If no resonance peak is observed in the reflected signal at cryogenic temperatures, the cavity length of the FBC can be readjusted by warming the system to room temperature and tuning the piezo mounting screws to realign the cavity as needed. The operating voltage of the piezoelectric actuator is actively controlled by the resonance reflection signal from a stabilized reference laser at 606 nm (Toptica, TA-SHG), which serves as the error signal. We reduced the power of the reference laser to 1 \textmu W, which virtually eliminated heating of the sample by the 606 nm laser, as evidenced by the measured optical coherence lifetime of 600 \textmu s for Eu$^{3+}$ ions. To achieve a prominent error signal, the reflectivity of both mirrors for the 606-nm reference light is 99\%. Using a side-of-fringe locking scheme, the cavity length is stabilized to the midpoint of the minimum resonance dip. As shown in Supplementary Figures \ref{fig:fiber cavity}c, the dual-wavelength resonance of the FBC at 580 nm and 606 nm is achieved by fine-tuning the wavelength of the 606-nm reference light across a broad range. Supplementary Figure~\ref{fig:fiber cavity}d shows a typical error-signal jitter after stabilization, measured at a working temperature of 4.3 K achieved with a stronger copper-strip thermal link. For actual experiments performed at 4.7 K, a softer thermal connection is employed, which further reduces the jitter to about half of that in Supplementary Figure~\ref{fig:fiber cavity}d. This corresponds to a cavity length fluctuation of roughly 50 pm, or about 1/64 of the linewidth under impedance-matching conditions.

Compared with high-finesse FBCs used to address single atoms in the Purcell regime \cite{doi:10.1126/sciadv.abo4538,Deshmukh:23} or in the strong coupling regime \cite{Niemietz2021NPQD,RevModPhys.94.041003}, where $\epsilon$ generally needs to be kept below 100 ppm, our FBC-based storage scheme imposes less stringent requirements on the fabrication and coating techniques of FBC. Owing to the relatively low finesse of 86 under impedance matching conditions, it is much easier to lock the cavity length within the resonance linewidth, which relaxes the demands on the vibration-isolation system.

\subsection*{6. Experimental setup of the FBC-based quantum memory}

As shown in Supplementary Figure~\ref{fig:FIG1}, the complete experimental setup for the FBC-based quantum memory consists of the quantum memory subsystem, the analysis subsystem, and the single-photon source subsystem. Two kinds of light sources are employed in experiments, i.e., weak coherent state pulses and telecom-heralded single photons. In the quantum memory part, the signal input beam and memory pump beam at 580 nm are combined into the same spatial mode. The reference light at 606 nm is combined with the 580-nm beams using a dichroic mirror (DM1). These beams are expanded by a factor of two, focused by a lens with a 100-mm focal length, and finally coupled into the FBC from the planar membrane side. The FBC is mounted on the vibration-isolation platform within a cryostat (Montana Instruments), with the temperature near the sample maintained at approximately 4.7 K. The membrane crystal and fiber concave mirror are bonded onto a sapphire substrate and a shear piezoelectric ceramic stack via epoxy resin, respectively. Input pulses propagate along the $b$-axis of the membrane with a polarization parallel to the $D_1$ axis of the $^{151}\mathrm {{Eu}^{3+}}$:$\mathrm {Y_2SiO_5}$ crystal. In the analysis part, we employ another spectrally-prepared $^{151}\mathrm {{Eu}^{3+}}$:$\mathrm {Y_2SiO_5}$ crystal to facilitate the analysis of superposition states of time-bin qubits encoded on weak coherent pulses.
The analyzer crystal is a 1\%-doped $^{151}\mathrm {{Eu}^{3+}}$:$\mathrm {Y_2SiO_5}$ crystal with a size of $5 \times 3 \times 20$ mm$^3$ along the crystal's $D_1 \times D_2 \times b$ axes located in another home-made cryostat with an operating temperature of 4.1 K. We implement a 520-ns AFC memory based on this analyzer crystal, ensuring that the intensity of transmitted pulses and retrieved AFC echoes are balanced. Since the 520-ns storage time corresponds to the time interval between the $\ket{e}$ and $\ket{l}$ qubits, the analyzer crystal effectively functions as an unbalanced Mach–Zehnder interferometer for measuring superposition states of time-bin qubits. After the analyzer crystal, the 580-nm signal is detected by a Si-SPD with a detection efficiency of about 69\% and a dark count rate of 30 Hz.

\begin{figure*}[tbph]
\includegraphics [width=1.0\textwidth]{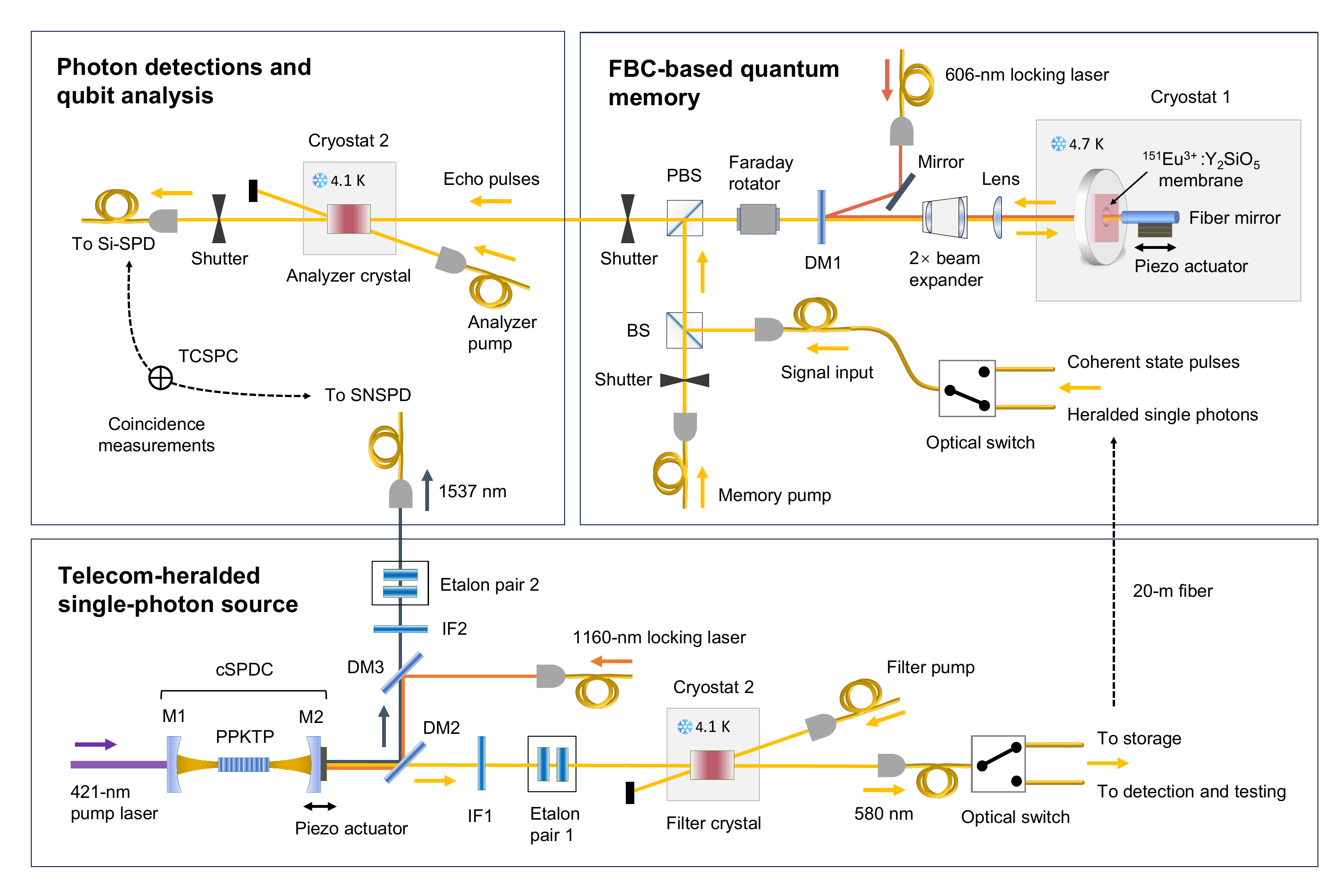}
\caption{Detailed experimental setup of the FBC-based quantum memory, the photon detection and qubit analysis system, and the telecom-heralded single photon source. Si-SPD, silicon-based single-photon detector; SNSPD, superconducting nanowire single-photon detector; TCSPC, time-correlated single-photon counting module; PBS, polarizing beam splitter; BS, beam splitter; DM, dichroic mirror; IF, interference filter; cSPDC, cavity-enhanced spontaneous parametric down-conversion; PPKTP, periodically poled potassium titanyl phosphate.}
\label{fig:FIG1}
\vspace{0.2cm}%
\end{figure*}

For quantum storage of weak coherent pulses, the coherent state pulses are attenuated to the single-photon level using neutral-density filters before being directed into the quantum memory. The read-out signal is separated from the input via a polarizing beam splitter (PBS) and a Faraday rotator. Following transmission through the analyzer crystal, the 580-nm weak pulses are detected by the Si-SPD and subsequently sent to a time-correlated single-photon counting module (TCSPC) for measurements triggered by electric signals. When the FBC is detuned from the resonance point, the signal detected by the Si-SPD is normalized by a factor of $R_{\mathrm{1(FBC)}}=96.5\%$, which serves as the reference for the input signals. The storage efficiency is then determined by comparing the retrieved AFC echoes from the FBC with the reference inputs. Since the signal light, after being transmitted through the coupling mirror, can still be reflected back into the original optical path by the fiber concave mirror, the actual input is smaller than the reference for the input signals we use here.  As a result, the storage efficiency reported here represents a lower bound of the actual efficiency.

The heralded single photons are generated through the cavity-enhanced spontaneous parametric down-conversion (cSPDC) process using a bulk periodically poled potassium titanyl phosphate (PPKTP) crystal~\cite{zhu2025metropolitan}, as shown in Supplementary Figure~\ref{fig:FIG1}. The PPKTP crystal, with dimensions of $0.8 \times 2 \times 20$ mm$^3$ and a quasi-phase-matching period of 4.775 \textmu m, is placed in a symmetric confocal cavity with an 18-cm cavity length. The 421-nm pump light is generated through sum-frequency generation from a frequency-stabilized 580-nm laser and a 1537-nm laser. The confocal cavity's input mirror (M1) and output mirror (M2) feature anti-reflection coating for 421 nm. For the generated 580-nm and 1537-nm photons, the input and output mirrors exhibit reflectivities of 99.8\% and 92\%, respectively, ensuring predominant photon emission through M2. 
We employ a frequency-stabilized 1160-nm laser as the locking beam, for which both M1 and M2 maintain 99\% reflectivity. The 580-nm laser is exactly aligned with the quantum memory's central absorption band, while two AOMs precisely tune the 1160-nm and 1537-nm wavelengths to achieve triple-wavelength resonance in the confocal cavity of cSPDC. The photon pairs generated by cSPDC are separated using a dichroic mirror (DM2) and subsequently filtered through interference filters (IFs) and cascaded etalons to ensure single-longitudinal-mode characteristics of the source. The etalon pair comprises two elements with linewidths of 3 GHz and 150 MHz, and free spectral ranges of 100 GHz and 2 GHz, respectively. The 1160-nm light is injected backward through another DM (DM3) in the 1537-nm path, implementing the Pound-Drever-Hall (PDH) locking for the cavity resonance. To precisely match the bandwidth of the quantum memory, the 580-nm photons are further filtered through a 1\%-doped $^{151}\mathrm {{Eu}^{3+}}$:$\mathrm {Y_2SiO_5}$ crystal filter (housed in the same cryostat as the analyzer crystal) with a size of $5 \times 3 \times 20$ mm$^3$, which has been prepared with a 4-MHz wide transmission window with near 100\% peak transmission through spectral hole-burning process. The final linewidths of the 580-nm and 1537-nm photons are 13 MHz and 4 MHz, respectively. 

For quantum storage of heralded single photons, the 1537-nm photon is detected by a superconducting nanowire single-photon detector (SNSPD) with a detection efficiency of about 85\% at a dark count rate of 10 Hz, and then sent to the TCSPC to perform coincidence measurements with 580-nm photons. The final source brightness is approximately 102 Hz in a 492-ns window before storage, at a pump power of 1 mW. The measurement of 580-nm photons is performed in a manner similar to that for weak coherent pulses, but without the involvement of the analyzer crystal. The measured storage efficiency is (69.8±1.6)\% for 2-\textmu s storage time. This efficiency already outperforms the fiber transmission at 580 nm, which is 33\% for a corresponding fiber length of 400 m, given a typical loss of 12 dB/km. The second-order cross-correlation measurements between 1537-nm photons and 580-nm photons yielded values of 20.7(2) and 16.4(2) before and after a 2-\textmu s storage, respectively, in a 123-ns coincidence window.

\subsection*{7. The storage fidelity for time-bin qubits}
The fidelity of the pole states $\ket{e}$ and $\ket{l}$ can be directly calculated using the formula $(S+N)/(S+2N)$. Here, $S$ and $N$ refer to the signal count and the noise count, respectively. The analysis of superposition states $\ket{e}+\ket{l}$ and $\ket{e}+\mathrm{i}\ket{l}$ requires an unbalanced Mach-Zehnder interferometer (MZI).

In the FBC, the superposition states $\ket{e}+\ket{l}$ and $\ket{e}+\mathrm{i}\ket{l}$ are encoded with two weak coherent Gaussian pulses with a FWHM of approximately 190 ns and a time spacing of 520 ns. To measure the fidelity of the retrieved superposition states, we employ another analyzer crystal prepared with a 520-ns AFC, which acts as an unbalanced MZI \cite{Duranti2024Efficient}. The transmitted part has the same intensity as the 520-ns AFC echo; therefore, there is an interference between the AFC echo of the $\ket{e}$ and the transmission of $\ket{l}$. The transmitted component has the same intensity as the 520-ns AFC echo, enabling good interference between the AFC echo of $\ket{e}$ and the transmission of $\ket{l}$. The fidelity of the superposition states can be determined from the interference visibility $V$ using the formula: $F_{+(-)}=(V+1)/2$.

In the WGC, we use two superimposed AFCs \cite{de2008solid-state} to measure the fidelity of the superposition states $\ket{e}+\ket{l}$ and $\ket{e}+\mathrm{i}\ket{l}$ \cite{Liu2020On-demand}. These qubits are encoded with two weak coherent Gaussian pulses with a FWHM of 260 ns and a spacing of 500 ns. For the analysis of the 1st echo, we create two combs with $1/\Delta=1$ \textmu s and 1.5 \textmu s, and optimize the finesse of AFCs to obtain balanced efficiencies. For the analysis of the 2nd AFC echo, the storage times of AFCs are set to 1 \textmu s and 1.25 \textmu s. After the electric-controlled rephasing, echoes are retrieved in 2 \textmu s and 2.5 \textmu s, enabling the interference between $\ket{e}$ stored for 2.5 \textmu s and $\ket{l}$ stored for 2 \textmu s. A variable phase shift of $2\pi\frac{\delta_f}{\Delta}$ between two AFCs is introduced by detuning the center frequency of AFCs, $\delta_f$, to measure the interference visibility \cite{Zhou2015Experimental}. 

By calculating the average fidelity of pole states, $F_{el}$, and the average fidelity of superposition states, $F_{+-}$, the total average fidelity $F_t$ can be expressed as: $F_t=\frac{1}{3}F_{el}+\frac{2}{3}F_{+-}$ \cite{Gundogan2015Solid}. Considering the Poisson distribution of coherent states, $P(\mu_q,m)=e^{-\mu_q}\mu_q^m/m$, an intercept-resend attack \cite{Marcos2005Intercept-resend} becomes feasible when the number of incident photons $m\geq m_{\mathrm{min}}$, where $m_{\mathrm{min}}$ is the minimal $i$ that satisfies $(1-P(\mu_q,0))\eta-\sum_{m>i+1}P(\mu_q,m)\geq 0$. The maximum fidelity achievable with a classical device can then be calculated as \cite{Gundogan2015Solid}: 
\begin{equation}
F_{\mathrm{classical}} = \frac{\frac{m_{\mathrm{min}}+1}{m_{\mathrm{min}}+2}\mathit{\Gamma}+\sum_{m\geq m_{\mathrm{min}}+1}\frac{m+1}{m+2}P(\mu_q,m)}{\mathit{\Gamma}+\sum_{m\geq m_{\mathrm{min}}+1}P(\mu_q,m)},
\end{equation}
where $\mathit{\Gamma}=(1-P(\mu_q,0))\eta-\sum_{m>m_{\mathrm{min}}+1}P(\mu_q,m)$. A storage fidelity, $F_t$, significantly exceeding $F_{\mathrm{classical}}$ provides conclusive evidence that the quantum memories are operating in the quantum regime.

All storage fidelities of time-bin qubits are summarized in Supplementary Table~\ref{TABLE 1}. The two fidelity-measurement schemes yield consistent results: the double-AFC method gives 99.6\% for the FBC, in agreement with the 99.4\% obtained using the AFC-based analyzer. Functionally, the two techniques are equivalent, though the latter physically separates the memory and analyzer stages. The AFC-based analyzer is employed only for the FBC, since an additional cryostat is required, and the crystal inside this cryostat conveniently serves both as the spectral filter for telecom-heralded single photons and as the analyzer.

To further investigate the phase coherence of temporally multiplexed storage, we measure the interference between the first AFC echo of the signal pulses and the residual higher-order AFC echoes of the reference input pulses. In the WGC, the reference input pulses are first incident on the memory, followed by the signal pulses after a delay of 1/$\Delta$ (3.8 \textmu s). This timing ensures the second echo of the reference input pulses interferes with the first echo of the signal pulses. Since the intensity of the residual second echo is much weaker than that of the first echo, the intensity of the reference input pulses is increased to balance these two echoes and maximize the interference visibility, as shown in the inset of Fig. 2d of the main text. Similarly, in the FBC, the residual third echo, which has the highest amplitude among the residual higher-order echoes, is selected as the reference to interfere with the first echo of the signal pulses, as illustrated in the inset of Fig. 3c in the main text. 

\begin{table*}

    \caption{\label{TABLE 1} \textbf{Storage fidelity of time-bin qubits.} 
     }
    \begin{ruledtabular}
    \begin{threeparttable}
    \begin{tabular}{ccccccccc}
        System &
        $\mu_q$ & 
        $\overline{\eta}$&
        $F_{|e\rangle}$ & 
        $F_{|l\rangle}$ & 
        $F_{|e\rangle+|l\rangle}$ & 
        $F_{|e\rangle+\mathrm{i}|l\rangle}$ & 
        $F_{T}$ & 
        $F_{\mathrm{Classical}}$
        \\ 
        \hline
        FBC &
        $0.7$ & 
        $73.2(5)\%$ &
        $99.9(1)\%$ &
        $99.5(1)\%$ & 
        $99.0(1)\%$ &
        $99.2(1)\%$ &
        $99.4(1)\%$ &
        $70.7\%$\\

        1st echo in WGC &
        $0.35$ & 
        $74.6(6)\%$ &
        $99.5(1)\%$ &
        $97.6(1)\%$ & 
        $99.2(1)\%$ &
        $99.1(1)\%$ &
        $99.0(1)\%$ &
        $68.6\%$\\

        2nd echo in WGC &
        $0.35$ & 
        $62.0(6)\%$ &
        $99.8(1)\%$ &
        $99.7(1)\%$ & 
        $99.6(1)\%$ &
        $99.2(1)\%$ &
        $99.5(1)\%$ &
        $69.0\%$
        \\
 
    \end{tabular}
     \begin{tablenotes}
        \footnotesize
        \item 
        $\overline{\eta}$: average storage efficiency. 
        $F_{|e\rangle}$: measured fidelity for $|e\rangle$.
        $F_{|l\rangle}$: measured fidelity for $|l\rangle$.
        $F_{|e\rangle+|l\rangle}$: measured fidelity for ${|e\rangle+|l\rangle}$.
        $F_{|e\rangle+\mathrm{i}|l\rangle}$: measured fidelity for ${|e\rangle+\mathrm{i}|l\rangle}$.
        $F_{T}$: total storage fidelity.
        $F_{\mathrm{Classical}}$: the maximal fidelity that can be achieved with the classical measured-and-prepare strategy. The slightly lower fidelity observed for $F_{|l\rangle}$ in the 1st echo in the WGC is attributed to imperfections in the AFC.
      \end{tablenotes}
  \end{threeparttable}
   
    \end{ruledtabular}

\end{table*}

\subsection*{8. Simulation of cavity-enhanced AFC memories}
The simulation of the quantum storage process is essential for optimizing various experimental parameters. In this section, we provide detailed results from these simulations. Here we employ a simplified model for the two-level AFC memory, treating the memory as a stable, dispersive medium with spectral modulation. Then, the input and the output of the electric field obey a simple relation:
\begin{equation}
\begin{aligned}
\mathcal{E}_{\mathrm{out}}(\nu)=\mathcal{E}_{\mathrm{in}}(\nu)e^{-ik^{\prime}(\nu)L},
\end{aligned}
\label{Eout bulk}
\end{equation}
where $L$ is the length of the memory crystal, $\nu$ is the frequency of the electric field. Therefore, we can simulate a two-level AFC memory using the complex wave vector $k^{\prime}(\nu)$. Under the first-order approximation, we express the wave vector as:
 \begin{equation}
\begin{aligned}
k^{\prime}(\nu)=k+\frac{k}{2n_0^2}(\chi^\prime(\nu)-i\chi^{\prime\prime}(\nu)),
\end{aligned}
\label{k vector}
\end{equation}
where $k$ and $n_0$ are the wave vector and refractive index in the crystal without absorption, respectively. $\chi^\prime(\nu)$ and $\chi^{\prime\prime}(\nu)$ are the real and imaginary parts of the susceptibility $\chi(\nu)$, which are related by the Kramers–Kronig relation:
\begin{equation}
\begin{aligned}
& \chi^\prime(\nu)=\operatorname{Re}(\chi(\nu))=\frac{2}{\pi} \int_0^{\infty} \frac{\nu^{\prime} \operatorname{Im}\left(\chi\left(\nu^{\prime}\right)\right)}{\nu^{\prime 2}-\nu^2} \mathrm{~d} \nu^{\prime}, \\
& \chi^{\prime\prime}(\nu)=\operatorname{Im}(\chi(\nu))=\frac{2}{\pi} \int_0^{\infty} \frac{\nu^{\prime} \operatorname{Re}\left(\chi\left(\nu^{\prime}\right)\right)}{\nu^{\prime 2}-\nu^2} \mathrm{~d} \nu^{\prime} .
\end{aligned}
\label{kk relation}
\end{equation}
Experimentally, the absorption profile $\alpha(\nu)$ can be measured, which is proportional to $\chi^{\prime\prime}(\nu)$ by the relation $\mathrm{Im}(k^\prime(\nu))=\alpha(\nu)$. By using Eq. \ref{k vector} and Eq. \ref{kk relation}, we can obtain the complex wave vector of the system.

\begin{figure}[tbph]
	\includegraphics [width=1.0\textwidth]{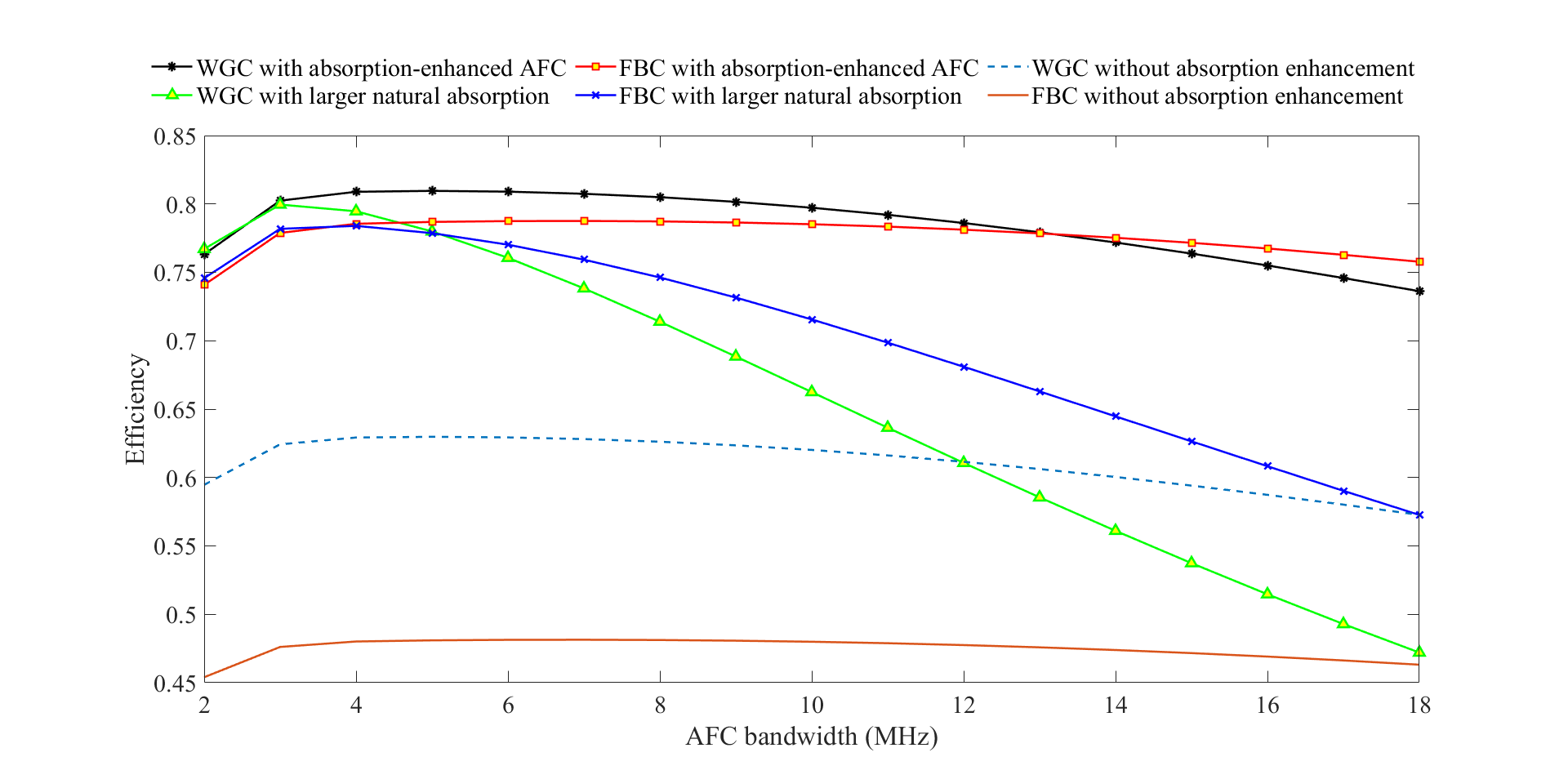}
	\caption{Simulated memory efficiency as a function of AFC bandwidth, maintaining the bandwidth of 20 MHz for the hole-burning pit. The linewidth of the input signal is set to half of the AFC bandwidth. We compare the memory efficiency of our absorption-enhanced AFC scheme with two assumed conditions: (i) memories using the same crystals in our experiments but without absorption enhancement, and (ii) memories using crystals with larger initial absorption, which is the same as the absorption after enhancement in our experiment.}
	\label{fig:eff vs AFCbw}
\end{figure}

For cavity-enhanced AFC memory, Eq. \ref{Eout bulk} should be replaced by \cite{POLLNAU2020100255}:
\begin{equation}
\begin{aligned}
\mathcal{E}_{\mathrm{out}}(\nu)=\mathcal{E}_{\mathrm{in}} (\nu)\frac{-\sqrt{R_1}+\sqrt{R_2} e^{-ik^{\prime}(\nu)L}}{1-\sqrt{R_1 R_2} e^{-ik^{\prime}(\nu)L}}.
\end{aligned}
\label{Eout cavity}
\end{equation}
For a realistic cavity with finite single-pass propagation loss $d_{\mathrm{loss}}$, $R_2$ should be replaced by $R_2^\prime=R_2e^{-2d_{\mathrm{loss}}}$. After performing a Fourier transform, we can obtain the time-domain retrieved electric field $\mathcal{E}_{\mathrm{out}}(t)$, and the echo signal can be written as:
\begin{equation}
\begin{aligned}
I_{\mathrm{echo}}(t)=\eta_\mathrm{M}\left\vert\mathcal{E}_{\mathrm{out}}(t)\right\vert^2.
\end{aligned}
\label{Eout cavity}
\end{equation}

\begin{figure}[tbph]
	\includegraphics [width=1.0\textwidth]{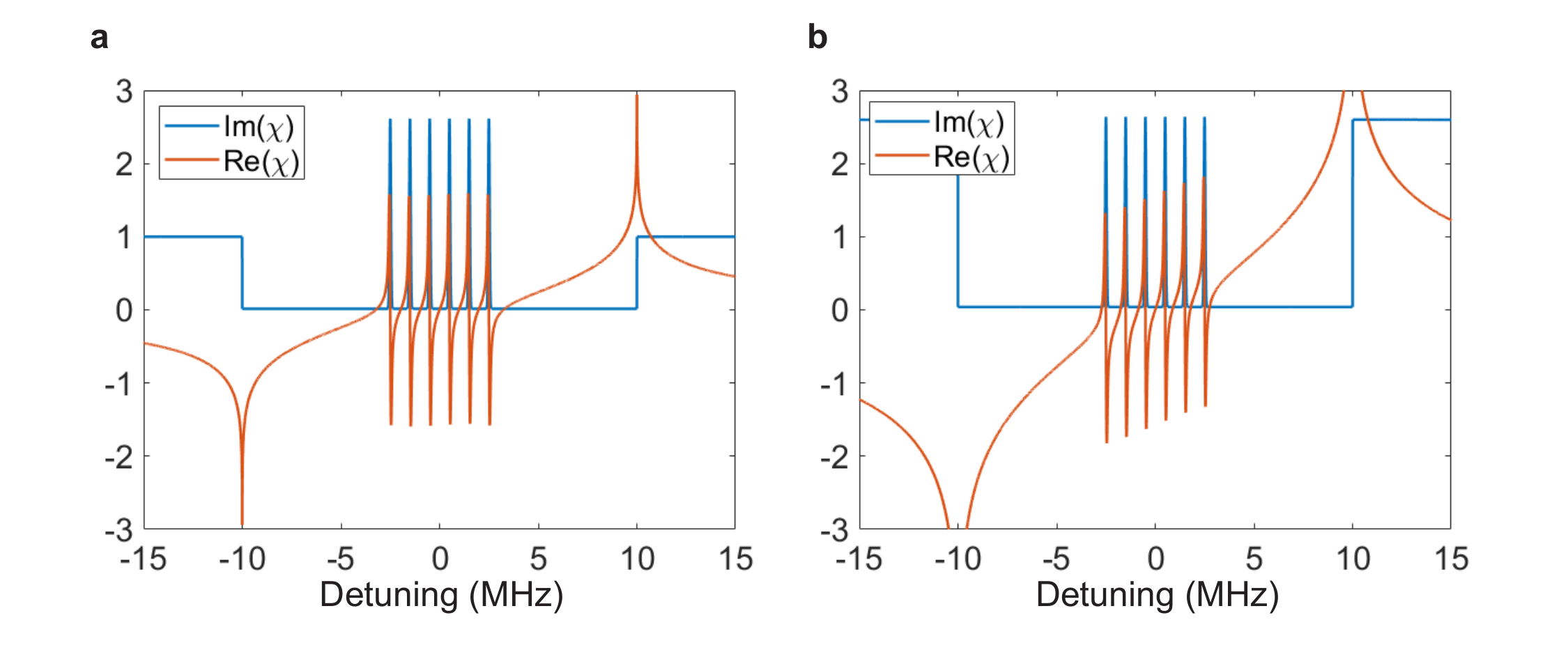}
	\caption{\textbf{a}. Real (red) and imaginary (blue) parts of the susceptibility $\chi(\nu)$ of a 5-MHz absorption-enhanced AFC sitting in a 20-MHz spectral pit. \textbf{b}.  Corresponding curves for a 5-MHz AFC prepared in a crystal with higher initial absorption. The absorption is set as the peak absorption in \textbf{a} (see both sides of $\mathrm{Im}(\chi(\nu))$).}
	\label{fig:dispersion}
\end{figure}

Using Eq. \ref{Eout cavity}, we simulated the echo trace presented in the main text. The input for the simulation was the fitted Gaussian function derived from the input signal. The theoretical efficiency can be calculated from the ratio of the areas under the echo and the input, consistent with the prediction derived from Eq. 1 in the main text. Based on Eq. \ref{Eout cavity}, we further evaluated the efficiency reduction of the FBC due to cavity length jitter. When the cavity length shifts from resonance by 50 pm (from $m\lambda/2$ to $m\lambda/2+50$ pm, where $m \in \mathbb{N}$ and $m\lambda\approx 460$ \textmu m), the simulated $\int\left\vert\mathcal{E}_{\mathrm{out}}(t)\right\vert^2/\left\vert\mathcal{E}_{\mathrm{in}}(t)\right\vert^2\mathrm{d}t$ decreases from 82.15\% to 82.02\%. In addition, $\eta_\mathrm{M}$ decreases from 95\% to 94.91\%, obtained by converting cavity length shift to resonance dip voltage similarly as shown in Supplementary Figure \ref{fig:fiber cavity}d. Overall, a 50-pm deviation in cavity length leads to a total efficiency reduction of approximately 0.2\%. Since cavity fluctuations are confined near the resonant point, the actual efficiency loss due to cavity instability is less than 0.2\%.

Our simulations clearly demonstrate the benefits of cavity linewidth and storage efficiency when adopting the absorption enhancement approach for preparing AFC with enhanced absorption. As illustrated in Supplementary Figure~\ref{fig:eff vs AFCbw}, applying the absorption enhancement scheme results in 18\% (WGC) and 33\% (FBC) increases in memory efficiency compared to using the same crystal without absorption enhancement, due to higher AFC finesse. 

Furthermore, compared to crystals with higher natural absorption (such as $\mathrm{Pr^{3+}:Y_2SiO_5}$), the absorption-enhanced AFC scheme presents a larger capacity of AFC bandwidth. As shown in Supplementary Figure~\ref{fig:eff vs AFCbw}, the efficiency of memories with absorption-enhanced AFC maintains almost no decrease when the AFC bandwidth becomes larger, when the efficiency using crystals with higher natural absorption declines significantly by contrast, which has also been observed in previous experiments~\cite{Duranti2024Efficient}.
That is because the fast light effect introduced by the AFC profile can neutralize the slow light effect caused by spectral hole burning, which was previously ignored in the analysis by Ref. \cite{Sabooni_2013}. The dispersion caused by spectral hole-burning is proportional to the ratio of absorption outside the spectral pit $d_{\mathrm{pit}}$ to spectral pit width $\Gamma_{\mathrm{pit}}$. On the other hand, the negative dispersion caused by the AFC profile is also proportional to the ratio of effective absorption $\widetilde{d}$ to AFC bandwidth $\Gamma_{\mathrm{AFC}}$. When the relation of $\frac{\widetilde{d}}{d_{\mathrm{pit}}}=\frac{\Gamma_{\mathrm{AFC}}}{\Gamma_{\mathrm{pit}}}$ is satisfied, the system bandwidth would be only limited by $\Gamma_{\mathrm{AFC}}$ rather than the slow-light cavity linewidth \cite{Sabooni_2013}. With the absorption enhanced AFC, the ratio $\frac{\widetilde{d}}{d_{\mathrm{pit}}}$ becomes higher, so this condition is satisfied with a larger ${\Gamma_{\mathrm{AFC}}}$ of 4.3 MHz (WGC) and 6.5 MHz (FBC), therefore providing a high temporal multimode storage capacity.

This mitigation of the dispersion can be easily understood by observing the shape of the susceptibility profile $\chi(\nu)$. Supplementary Figure~\ref{fig:dispersion} shows the susceptibility of absorption-enhanced AFC compared to crystals with higher natural absorption. When a 5 MHz absorption-enhanced AFC is created in a 20 MHz spectral pit, the real part of the susceptibility exhibits a perfect periodicity within the AFC bandwidth, representing an AFC memory with no dispersion. For the crystals with higher natural absorption but no enhancement, the real part of susceptibility shows an overall slope in the AFC bandwidth, which suggests the detrimental slow-light effects.

\begin{figure}[tbph]
	\includegraphics [width=1.0\textwidth]{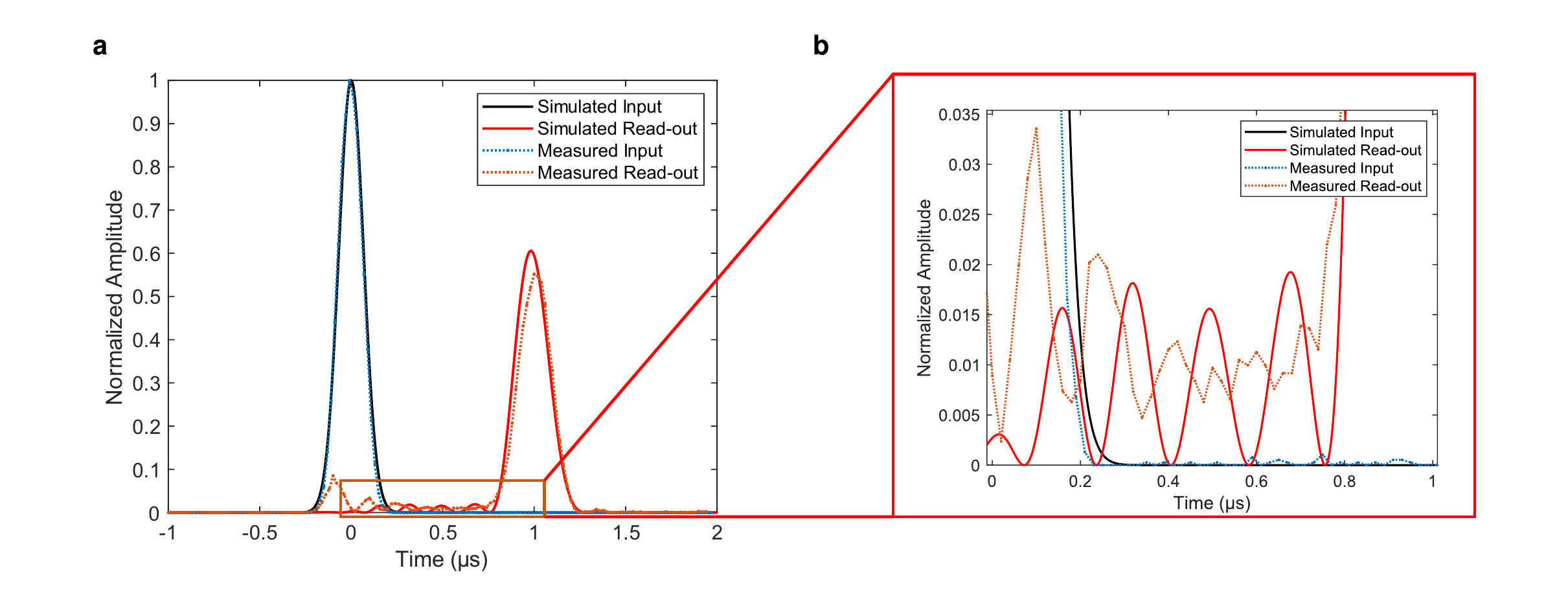}
	\caption{Simulations and experimental results of the storage of time-bin qubit $\ket{l}$ in the WGC. Here, we assume only five teeth of the AFC are successfully created. Some noise peaks appear before the retrieved echo, which are observed in both our simulations and experiments.}
	\label{fig:WGC qubit noise}
\end{figure}

In the WGC, the cavity linewidth is slightly narrower than the desired AFC bandwidth, complicating the precise tailoring of the outermost AFC teeth. The absence of these teeth introduces noise before the retrieved echo when the input signal has a large bandwidth. As shown in Supplementary Figure~\ref{fig:WGC qubit noise}, storing a short pulse in the WGC leads to some noise peaks appearing before the main echo. These noise peaks could reduce the fidelity of the $\ket{l}$ qubit since they are interpreted as noise during the analysis of qubits (see Supplementary Table \ref{TABLE 1}). 

It is important to note that the current simulation based on spectral analysis is exclusively applicable to the storage of optical pulses in two-level AFC. The dynamical storage process with electric control and the storage of a single-photon wave-packet exceed the capabilities of this approach. A more accurate model with complete quantum treatment for both the light and atoms, capable of analytically describing the entire process of cavity-enhanced quantum memories, will be presented in a forthcoming work.

\end{document}